\documentclass[showpacs,preprintnumbers,twocolumn,amsmath,amssymb,
prd,a4paper,floatfix,nofootinbib]{revtex4-1}

\usepackage[bookmarks=false]{hyperref}

\usepackage{graphics}
\usepackage{graphicx}
\usepackage{epsfig}
\usepackage{color}
\usepackage{longtable}
\usepackage{hyperref}
\usepackage{latexsym}
\usepackage{amsmath}
\usepackage{amsthm}
\usepackage{amsfonts}
\usepackage{amssymb}
\usepackage{dsfont}
\usepackage{bm}
\usepackage{graphics,psfrag}
\usepackage{graphicx,psfrag}
\usepackage[nice]{nicefrac}
\newcommand{\ov}[1]{\overline{#1}}
\newcommand{\be}{\begin{equation}}
\newcommand{\ee}{\end{equation}}
\newcommand{\bea}{\begin{eqnarray}}
\newcommand{\eea}{\end{eqnarray}}
\newcommand{\eq}[1]{(\ref{#1})}

\newcommand{\DCI}{\ensuremath{D_\srm{CI}} }

\newcommand{\VEC}[1]{\ensuremath{\textrm{\boldmath{#1}}}}
\newcommand{\la}{\langle}
\newcommand{\ra}{\rangle}

\newcommand{\srm}[1]{\textrm{\scriptsize{#1}}}

\newcommand{\E}{\mathrm{e}}
\newcommand{\FC}{\;,}
\newcommand{\FD}{\;.}
\renewcommand{\imath}{\mathrm{i}}
\newcommand{\dbar}{\overline{d}}

\newcommand{\myparagraph}[1]{~\\ \noindent\ensuremath{\mathbf{#1}}:~}

\begin{document}
%
\title{QCD with two light dynamical chirally improved quarks: Mesons}
\author{Georg P.~Engel$^1$, 
C.~B.~Lang$^1$, 
Markus Limmer$^1$, 
Daniel Mohler$^{1,2}$, and
Andreas Sch\"afer$^3$\\
\vspace*{2mm}
(BGR [Bern-Graz-Regensburg] Collaboration)
\vspace*{2mm}\\}

\affiliation{
$^1$Institut f\"ur Physik, FB Theoretische Physik, Universit\"at
Graz, A--8010 Graz, Austria\\
$^2$TRIUMF, 4004 Wesbrook Mall Vancouver, BC V6T 2A3, Canada \\
$^3$Institut f\"ur Theoretische Physik, Universit\"at
Regensburg, D--93040 Regensburg, Germany
}

\date{\today}

\begin{abstract}
We present results for the spectrum of light and strange mesons on
configurations with two flavors of mass-degenerate Chirally Improved sea quarks.
The calculations are performed on seven ensembles of lattice size $16^3\times 32$
at three different gauge couplings and with pion masses ranging from 250 to 600
MeV. To reliably extract excited states, we use the variational method with an
interpolator basis containing both gaussian and derivative quark sources. Both
conventional and exotic channels up to spin 2 are considered. Strange quarks are
treated within the partially quenched approximation. For kaons we investigate
the mixing of interpolating fields corresponding to definite C-parity in the
SU(3) limit. This enlarged basis allows for an improved determination of the
low-lying kaon spectrum. In addition to masses we also extract the ratio of the
pseudoscalar decay constants of the kaon and pion and obtain
$F_K/F_\pi=1.215(41)$. The results presented here include some ensembles from
previous publications and the corresponding results supersede the previously
published values.
\end{abstract}


\maketitle

\section{Introduction}

Considering only strong decays, with the exception of the pion and the proton all hadrons are resonances,
embedded in a continuous spectrum. In lattice calculations we can only
determine discrete energy levels, with spacings 
$\mathcal{O}(1/L)$ 
related to the spatial extent $L$ of the studied lattice volume. 
When disregarding the
fermion vacuum in the so-called quenched simulations energy levels can be
related directly to hadron excitations. In dynamical situations the energy
levels are denser close to resonances and they are influenced by
coupled open hadronic scattering channels. Although in principle the
Euclidean correlator of any hadron interpolator with the correct quantum
numbers should feel these scattering channels, in actual calculations there is
little, if any, trace of it \cite{Engel:2010my,Dudek:2010wm}  unless such
multi-hadron interpolators are included explicitly in the set of operators.
However, inclusion of those is costly, since it involves disconnected
contributions. 
In actual calculations efficient but demanding all-to-all propagator methods are used  
\cite{Foley:2005ac,Peardon:2009gh,Morningstar:2011ka,Bali:2009hu,Bali:2010se}.

In recent years much effort has been invested into developing methods for
determining the lowest energy levels for hadron correlators. In
\cite{Petry:2008rt,Burch:2009wu,Fleming:2009wb,Dudek:2009qf,Engel:2010my,Dudek:2010wm,Dudek:2011tt,Morningstar:2011ws}
meson excitations have been studied in a dynamical quark background with a
variety of quarks species, interpolators and extraction methods. A central
technique employed was the variational method
\cite{Luscher:1990ck,Michael:1985ne} where one finds the energy levels by
diagonalization of cross-correlations of a (hopefully) sufficiently large set 
of interpolators which allows for a good overlap with the relevant hadron states.

In continuum quantum field theory there has been
recent progress in investigations of mesons 
using Schwinger-Dyson equations and the Bethe-Salpeter equation 
as well as effective field theories (see for examples Refs.
\cite{Krassnigg:2009zh,Blank:2010pa,Krassnigg:2010mh,Pelaez:2010fj,
Bernard:2010fp,Doring:2011vk,Doring:2011nd}).

Starting with \cite{Gattringer:2008vj} we have been determining hadron ground
states and low  excited states in a framework of simulations with two light
dynamical quarks. The fermionic action used is the so-called Chirally Improved
(CI) action \cite{Gattringer:2000js,Gattringer:2000qu}, an approximate
solution to the Ginsparg-Wilson relation for fermions obeying chiral  symmetry
in a lattice form. The strange quarks have been incorporated in the valence
sector only. In \cite{Engel:2010my} results based on three ensembles at three
different gauge couplings but with only one quark mass for each coupling have been 
presented. 
We have meanwhile significantly extended the statistics and also the number
of ensembles. Here, we  present our results for the meson sector based on the
final set of seven ensembles at three gauge couplings and two or three quark
mass values at each. This allows an extrapolation  towards the physical
point. Previously published results are generally confirmed, although
in  some cases we observe new behavior related to new symmetry considerations.
Some results have been presented already in \cite{Engel:2011pp}.

Following the presentation of the action and the parameters of the gauge
configuration ensembles  in Sec. \ref{action} we discuss scale setting, decay
constants and the quark mass in Sec.  \ref{scale}.  The interpolators used for
the meson fields in the variational analysis are discussed in Sec.
\ref{analysismethod} and tabulated in the appendix.  The main parts are Secs
\ref{lightmesons} and \ref{strangemesons},  where results for the mesons are 
presented.  

\section{Action and simulation}\label{action}

\subsection{Fermion action and gauge action}

In our study the fermions are represented by the Chirally Improved Dirac operator \DCI 
\cite{Gattringer:2000js,Gattringer:2000qu}.  
This is an approximate solution of the Ginsparg-Wilson equation
and results from a general ansatz for the Dirac operator, namely an
expansion of the form
\be
D= m_0 \mathds{1} + \DCI\ , \ \DCI(n,m)= \sum_{i=1}^{16} c_{nm}^{(i)}(U)\; \Gamma_i\ ,
\ee
where the sum runs over all $16$ elements $\Gamma_i$ of the Clifford algebra 
and the coefficients $c^{(i)}_{nm}$ 
were fit by minimizing the violation of the Ginsparg-Wilson equation.
It includes paths up to a maximum length of 4 lattice units.  The  paths
and coefficients used  are found in the appendix of \cite{Gattringer:2008vj}.
We used the same 19 coefficients for all ensembles, modifying only the
diagonal mass term in order to account for the additive mass renormalization.
For that reason the values of the bare mass parameter $m_0$ given in Table
\ref{tab:ensembles} are negative. Thus the actual (unrenormalized) mass is
given  by the values $m_{AWI}$ determined from the axial Ward identity. 

For further improvement of the fermion action one level of stout smearing of
the gauge fields  \cite{Morningstar:2003gk} was included in its definition. The
parameters are adjusted such that the value of the plaquette
is maximized ($\rho=0.165$ following \cite{Morningstar:2003gk}). For the pure
gauge field part of the action we use the tadpole-improved L\"uscher-Weisz 
gauge action \cite{Luscher:1984xn}. For a given gauge coupling
we used the same assumed plaquette value for the different values of the bare
quark mass parameter.

\subsection{Lattice ensembles}

The analysis presented here is based on seven ensembles of configurations  for
lattice size $16^3\times 32$. These substantially extend (by a factor of
three) the data base of \cite{Gattringer:2008vj,Engel:2010my}. A summary of the
notation and some parameters of these ensembles is given in Table
\ref{tab:ensembles}. 

The notation for the couplings follows \cite{Gattringer:2008vj}, where all
parameters of the fermion action are detailed. For each value of the gauge
coupling we have two or three values of the quark mass parameter.  Following
equilibration every 5th configuration has been selected for analysis. Further
details on the updating HMC-method and statistical checks for equilibration have
been discussed in \cite{Gattringer:2008vj}.

From the values of $m_\pi L$ we expect non-negligible finite size effects for the three
ensembles with smallest quark mass, A66, B70 and C77.
Discretization effects have been discussed in the quenched
simulations, where for the used action only small
$\mathcal{O}(a^2)$ corrections have been identified \cite{Gattringer:2003qx}.
In order to confirm this for the dynamical simulation we would have to perform
our study at several lattice spacings and volumes, which is
not possible based on the given ensembles and statistics.
Studies with a larger volume ($24^3\times 48$) 
with linear size $\mathcal{O}(3.6\;\textrm{fm})$ are in progress.

\begin{table}[tbp]
\begin{ruledtabular}
\begin{tabular}{cccccc}
set&	$\beta_{LW}$	&$m_0$	&$m_s$& configs 	&$m_{\pi}L$ \\
\hline
A50&	4.70& 		 -0.050	&-0.020	&200 	& 6.4\\
A66&	4.70& 		 -0.066	&-0.012	&200 	& 2.7\\
B60&	4.65& 		 -0.060	&-0.015	&300 	& 5.7\\
B70&	4.65& 		 -0.070	&-0.011	&200 	& 3.4\\
C64&	4.58& 		 -0.064	&-0.020	&200 	& 6.7\\
C72&	4.58& 		 -0.072	&-0.019	&200 	& 5.1\\
C77&	4.58& 		 -0.077	&-0.022	&300 	& 3.7
\end{tabular}
\end{ruledtabular}
\caption{\label{tab:ensembles}Parameters of the simulation: 
We used several ensembles with different gauge couplings
$\beta_{LW}$ and/or light quark mass parameters $m=0$.
We also show the strange quark mass parameter $m_s$, the number of
configurations analyzed and the physical extent of the spatial volume
multiplied with the pion mass.}
\end{table}

\section{Scale and low energy parameters}\label{scale}
\subsection{Scale}\label{subsec:scale_light}

In our earlier work \cite{Gattringer:2008vj,Engel:2010my} we had analyzed 
configurations at one quark mass parameter for three values of the
gauge coupling. There, we used the lattice spacing derived from the
static potential with a Sommer parameter $r_0=0.48$ fm. Now we have two or
three quark mass parameters for each gauge coupling and can attempt an extrapolation to the physical
point or the chiral limit. The latter extrapolation would be relevant for the
parameters of Chiral Perturbation  Theory (ChPT), which we will not attempt to extract
here.

\begin{figure}[tbp]
\centering
\includegraphics[width=\columnwidth,clip]{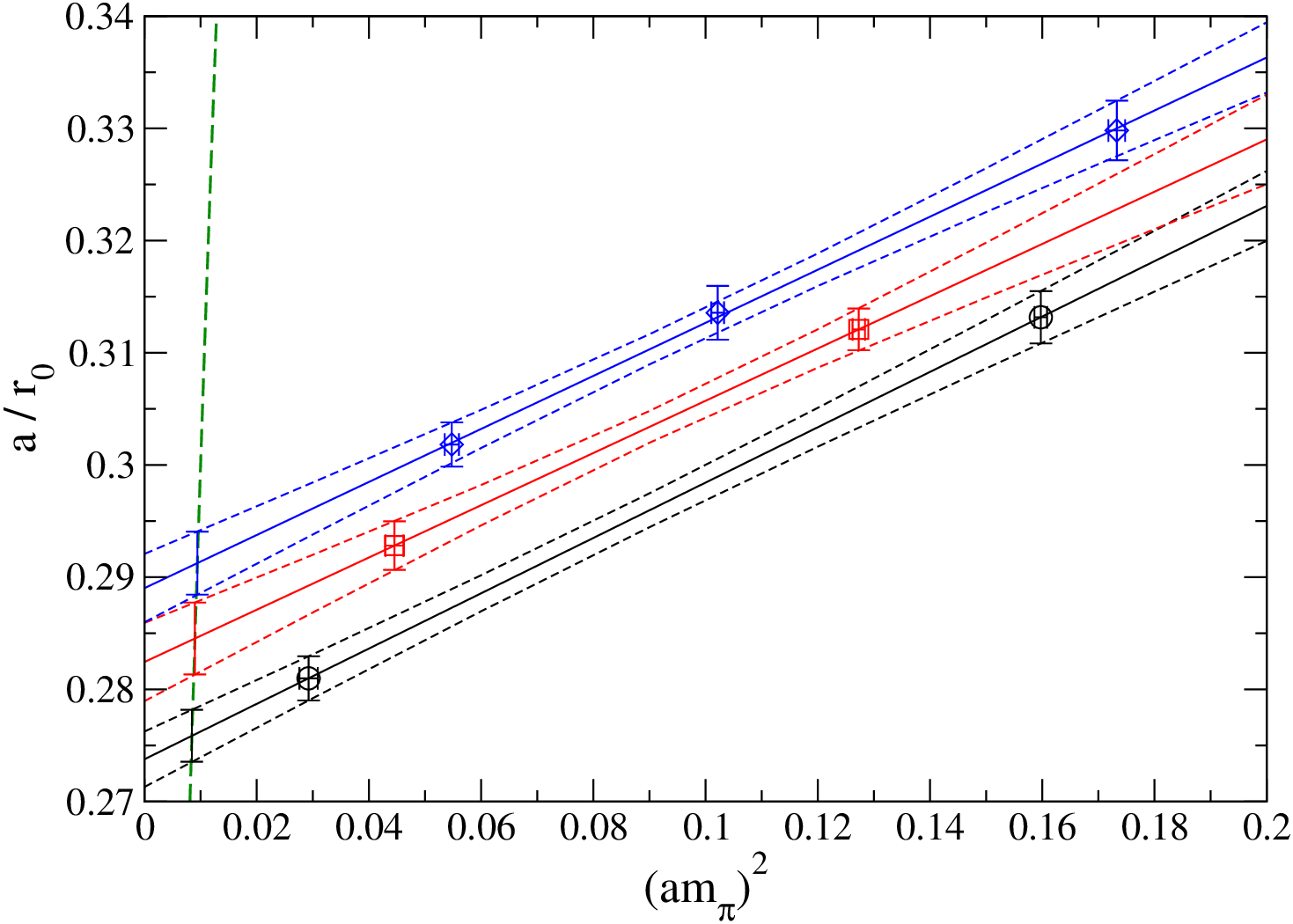}
\caption{(color online). Setting the scale with the Sommer parameter and the pion mass as input at
the physical point.
The green (long-dashed) line is the curve Eq.~(\ref{ar0_vs_ampisq}). The solid
and short-dashed lines represent the extrapolation of our lattice data.
Their intersections with the green line define the lattice constants $a$.
}
\label{fig_scale_r0}
\end{figure}

We use two approaches to set the scale. In the first one we determine $y\equiv
a/r_0$ from the static potential separately for each ensemble, as discussed in
\cite{Gattringer:2008vj}. We then study the dependence of this quantity on the
measured values of $x\equiv(a m_\pi)^2$ (cf., Fig. \ref{fig_scale_r0}). The
physical values are obtained along
\be\label{ar0_vs_ampisq}
y=\frac{\sqrt{x}}{m_\pi r_0}\FD
\ee
For each of the three gauge couplings we then perform a linear fit and obtain
the physical value where the extrapolations intersect Eq.~\eq{ar0_vs_ampisq} with 
$m_\pi r_0 = 137 \,\textrm{MeV} \times 0.48\, \textrm{fm} = 0.3332$. 
(We use the average of charged and neutral pion masses.)
From this one
reads off the lattice spacing $a$. Table \ref{tab_scale} gives the resulting
value in the row labeled $(\pi,r_0)_{phys}$. 
The value in the chiral limit is obtained as usual from $a/r_0$ where
$a m_\pi=0$.

\begin{table}
\begin{ruledtabular}
\begin{tabular}{lccc}
			& A	      & B          &C         \\
\hline
$(\pi,r_0)_{phys}$	& 0.1324(11)  &0.1366(15)  &0.1398(14) \\
$(\pi,r_0)_{chiral}$	& 0.1314(12)  &0.1356(17)  &0.1387(15) \\
$(\pi,\rho)_{phys}$	& 0.1330(44)  &0.1378(50)  &0.1400(29) 
\end{tabular}
\end{ruledtabular}
\caption{\label{tab_scale}Lattice spacing in physical units derived for
ensembles of type A, B, C (cf., Table \ref{tab:ensembles}) by the methods
discussed in the text.} 
\end{table}

The other approach is to replace $y=a/r_0$ by mass values like $a m_N$ or $a
m_\rho$. Since the  $\rho$ is unstable for small enough pion mass, there will
be threshold effects. In our parameter range we find no coupling  to the
(p-wave) $\pi\pi$ sector yet and a linear extrapolation intersecting with
$y=\sqrt{x}\, m_\rho/m_\pi$ gives the values of the lattice spacing in Table
\ref{tab_scale} compatible with the results of the first method, but with larger errors. 

Throughout this presentation we will use the values obtained from the definition
denoted by $(\pi,r_0)_{phys}$ in Table \ref{tab_scale}.

\subsection{Setting the strange quark mass}\label{subsec:scale_strange}
In this two-flavor simulation we use the partial quenching approximation to
access the strange hadron spectrum,i.e., we consider the strange quark as a valence quark only.
In view of results with full strange quark dynamics (e.g., 
\cite{Durr:2008zz}) we find,  at least for the ground states, no 
noticeable difference in the mass range considered here.
In each ensemble the strange quark mass
parameter $m_s$ is set by identifying our result  for the $\Omega$ baryon positive
parity ground state energy level with the physical $\Omega(1672)$. These
parameters are found in Table \ref{tab:ensembles}. 

For this definition we use $r_{0,exp}=0.48\,$fm in each ensemble,
differing from the (in Subsec. \ref{subsec:scale_light}) 
discussed method to set the overall scale.
Since the two different definitions agree at physical pion masses,
this method is consistent at the physical
point, but results have to be taken with care at unphysically large pion masses.

\subsection{AWI mass}

The so-called axial Ward identity (AWI) mass (or PCAC mass) is determined from
the asymptotic (i.e., plateau of the) ratio of the unrenormalized correlators
\be \label{eq:awi_renorm}
2\,m_\srm{AWI} = \frac{c_A}{c_P}
 \frac{\la 0\vert\partial_t A_4^+(\VEC{p}=0,t)\ X(0)\vert 0\ra}
 {\la 0\vert P^-(\VEC{p}=0,t)\ X(0)\vert 0\ra}\FC
\ee
where $P^-=\dbar \gamma_5 u$, $A_4^-=\dbar \gamma_4\gamma_5 u$, and $X$ is an
interpolator with the quantum numbers of the pion, usually  $P^+$ or $A^+$. The
constants $c_A(s)$ and $c_P (s)$ denote the lattice factors relating the
smeared interpolators to the lattice pointlike interpolators (not to be
confused with the renormalization constants $Z$ relating lattice point
operators to the continuum renormalization scheme). They are obtained from the
ratio of correlators from smeared to point sources \cite{Gattringer:2008vj}.

The relation to the renormalized quark mass needs the renormalization factors
for the pseudoscalar and axial currents,
\be
m^{(r)} = \frac{Z_A}{Z_P}\, m_\srm{AWI}\FD
\ee
Table \ref{tab:masses} gives the values of $m_\srm{AWI}$ and $m_\pi$ for the
ensembles studies. 
(Values for the renormalization constants have been derived in 
\cite{huber:2010zza,Maurer:2011}.)

\begin{table}
\begin{ruledtabular}
\begin{tabular}{cccccc}
Set & $a$ & $a\, m_\pi$ &$m_\pi$ & $a\, m_\srm{AWI}$ & $m_\srm{AWI}$  \\
    &[fm] &             &   [MeV]&                   &          [MeV] \\
\hline
A50 & 0.1324(11)  & 0.3997(14) &596(5) &0.03027(8)  & 45(1)  \\
A66 & 0.1324(11)  & 0.1710(48) &255(7) &0.00589(40) &  9(1)  \\
B60 & 0.1366(15)  & 0.3568(15) &516(6) &0.02356(13) & 34(1)  \\
B70 & 0.1366(15)  & 0.2111(38) &305(6) &0.00836(23) & 12(1)  \\
C64 & 0.1398(14)  & 0.4163(18) &588(6) &0.02995(20) & 42(1)  \\
C72 & 0.1398(14)  & 0.3196(18) &451(5) &0.01728(16) & 24(1)  \\
C77 & 0.1398(14)  & 0.2340(27) &330(5) &0.01054(19) & 15(1) 
\end{tabular}
\end{ruledtabular}
\caption{Pion masses and quark AWI-masses for the different sets of gauge
configurations. }
\label{tab:masses}
\end{table}

\subsection{Decay constants}
The pseudoscalar decay constant describes the coupling to weak decays. It can
be extracted from the asymptotic behavior of the correlation between the
pseudoscalar or the time components of the axial interpolators.
\be \label{eq:def_fpi1}
c_A^2\, Z_A^2\, \la A_4^-(\VEC{p}=0,t)\, A_4^+(0) \ra \sim
m_\pi\, F_\pi^2\, \E^{-m_\pi t}\equiv c \,\E^{-m_\pi t}\FD
\ee
The coefficient then gives
\be
F_\pi =
2\, m_\text{AWI}\, c_P\, Z_A \sqrt{\frac{c}{m_\pi^3}}\FC
\ee
and equivalently for the kaon $F_K$.

The dependence of the pion decay constant on the quark mass can be described
by chiral perturbation theory. Up to $1$-loop order one finds
\cite{Gasser:1983yg}
\be \label{eq:fpi_from_chpt}
F_\pi = F_{\pi,0}-m\, \frac{2\,\Sigma_0}{16\,\pi^2 F_{\pi,0}^3}\,
\ln\left(m\frac{2\,\Sigma_0}{\Lambda_4^2 F_{\pi,0}^2}\right) \ .
\ee
Here, $F_{\pi,0}$ and $\Sigma_0$ refer to the pion decay constant and the
quark condensate in the chiral limit $m\to 0$ and $\Lambda_4$ is a low energy
constant. The corresponding expressions including the $2$-loop order can be
found in \cite{Colangelo:2001df,Bijnens:2007yd}.

\begin{figure}
\begin{center}
\includegraphics[width=\columnwidth,clip]{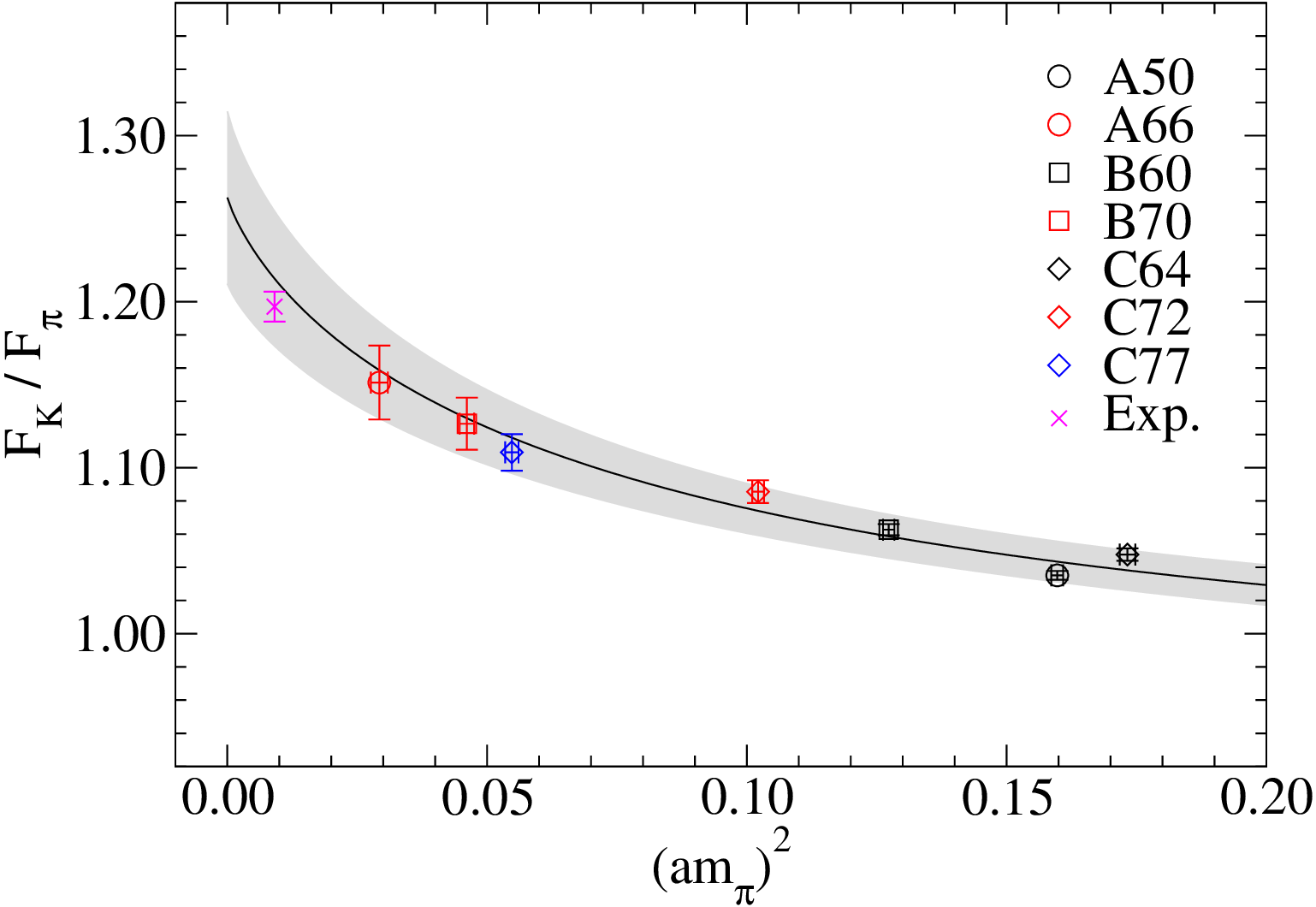}
\caption{The ratio of $F_K/F_\pi$ is plotted against $m_\pi^2$ (in
dimensionless units) for each set of gauge configurations. The full black line
is a fit of the data using the relevant expressions for numerator and
denominator; the shaded area indicates the error band. 
The magenta cross indicates the experimental value
\cite{Nakamura:2010zzi}.}
\label{fig:ratio_k_over_pi}
\end{center}
\end{figure}

The renormalization factor $Z_A$ cancels in the ratio $F_K/F_\pi$. We show
this ratio in Fig. \ref{fig:ratio_k_over_pi} where we assume a lattice spacing
of $0.135$ fm (the average of our values for the scheme $(\pi,r_0)_{phys}$) and a
physical pion mass of $139.57$ MeV. The extrapolation of our data to that
point gives
\be
F_K/F_\pi = 1.215(41)\FD
\ee
which fully covers the experimental value 1.197(9)\cite{Nakamura:2010zzi}.

\section{Analysis method and meson interpolators}\label{analysismethod}

Given interpolating operators $O_M$ with the quantum numbers of a hadron, the
correlation function of such operators separated by some Euclidean time
distance provides the energy spectrum,
\be
\langle O_M(t) O_M^\dagger(0)\rangle=\sum_n \langle O_M|n\rangle\langle n|O_M^\dagger\rangle
\E^{-E_n t}\FD
\ee
The asymptotic exponential decay, however, gives just the ground state energy
in that channel. On finite lattices, depending on parameters like size and
lattice spacing, this may be related either to a single meson or to meson scattering states. For the study of scattering and of higher lying
mesons it is imperative to find also the excited energy levels. 

An efficient tool for this is the so-called variational analysis
\cite{Luscher:1990ck,Michael:1985ne,Blossier:2009kd}. Using several
interpolators with the correct quantum numbers, one diagonalizes the
cross-correlation matrix of these, using the generalized eigenvalue
formulation
\bea
C_{ij}(t)&\equiv& \langle O_i(t) O_j^\dagger(0)\rangle\FC\nonumber\\
C(t) \vec v_k(t,t_0)&=&\lambda_k(t,t_0) C(t_0) \vec v_k(t,t_0)\FD
\eea
If the set of interpolators is large enough, then one expects that the
eigenvectors approach the eigenstates of the system. In fact, the eigenvectors
act as a fingerprint of the states and should remain stable over the
considered window of $t$-values. In such a window the eigenvalues decay
exponentially, approximating the desired eigenenergies,
\be
\lambda_k(t,t_0) \propto \E^{-(t-t_0)E_k}\left(1+\mathcal{O}(\E^{-(t-t_0)\Delta E_k}\right)\FD
\ee
Here, depending on $t$ and $t_0$ the value of $\Delta E_k$ denotes either the
difference to the first neglected energy level (for $t_0\le t \le 2 t_0$) or
to the nearest energy level (for a  careful discussion see
\cite{Blossier:2009kd}). It was also demonstrated, that even ghost states can
be identified with this type of analysis \cite{Burch:2005wd}.

A possible systematic influence comes from choosing $t_0$ in the
variational method and the fit range for the generalized eigenvalues. 
We use $t_0=1$ throughout. In
principle, the impact of that choice can be estimated by choosing several values of $t_0$ and
varying the fit range. For the final fit one should then choose a window where
this impact is negligible. However, in practice the corresponding choices are
restricted by the given signal-to-noise ratio for coarse lattices and weak
signals.
In the actual analysis one determines the window from a combination of
indicators, ranging from effective energy values to approximate constancy of the corresponding
eigenvectors. The energy levels then result from an exponential fit to the
eigenvalues over that window. In some cases a second exponential is used in these fits to allow for a small admixture of higher energy
states.

Various techniques have been suggested to construct interpolators. In
\cite{Burch:2004he} we introduced lattice operators based on smeared quarks.
Combining differently smeared quarks, also including covariant derivatives
\cite{Liao:2002rj,Dudek:2007wv,Gattringer:2008be}, several meson and baryon energy levels could be
determined in the quenched \cite{Burch:2006cc,Burch:2006dg} and dynamical case
\cite{Engel:2010my}.

The interpolators are constructed by hypercubic (HYP)-smearing
\cite{Hasenfratz:2001tw,Hasenfratz:2001hp,Hasenfratz:2007rf} the time slice
gauge variables, i.e., smearing only the spatial links in each time 
slice \footnote{Notice that the Dirac operator already contains one level of stout smearing. We use these stout smeared gauge links and apply additional smearing to construct the sources.}.
Based on these gauge variables the quark sources are smeared with the
covariant Jabobi smearing \cite{Gusken:1989ad,Best:1997qp} 
\begin{eqnarray}
S_{\kappa,K} 		&=& \sum_{n=0}^K\kappa^nH^n S_0\,,  \\
H(\vec{n},\vec{m})	&=& \sum_{j = 1}^3 \Big(U_j\left(\vec{n},0\right) 
\delta\left(\vec{n} + \hat{j}, \vec{m}\right) \vspace{-6pt}\\
&&  \phantom{\sum_{j = 1}^3}+ U_j\left(\vec{n}-\hat{j\,},0\right)^\dagger \delta\left(\vec{n} - \hat{j},
 \vec{m}\right) \Big)\,,\nonumber
\end{eqnarray}
where $S_0$ denotes the point source.  The parameters $K$ and $\kappa$ are
adjusted to obtain gaussian-like shapes of the sources \cite{Burch:2006dg} with
different smearing widths. In the definitions of the operators we denote the
smearing types by $n$ and $w$ (narrow and wide) and by $\partial_k$ for the
derivative in spatial direction $k$. 
The widths of the sources do not exactly agree for the various ensembles (which would be
dependent on the definition of the scale anyway.) However, the width of the narrow source is in the range 0.2 to 0.3
fm and the width of the wide source is in the range 0.4 to 0.6 fm. 

The derivative sources $S_{\partial_k}$
have been constructed numerically by applying the covariant difference operators on the wide
source, $S_w$, see \cite{Gattringer:2008be,Gattringer:2008td}.
This corresponds to an asymmetric definition of the interpolators.
If $S^1,S^2$ denote gaussian smearing operators and $\overrightarrow{D}$ the derivative acting to the right, then our operators (involving one derivative) have the structure
\be
O=\bar{\psi} (S^1 \Gamma S^2 \overrightarrow{D} \pm \overleftarrow{D}  S^2 \Gamma S^1) \psi \;
\label{eq:der_op1}
\ee
instead of
\be
O=\bar{\psi} (S^1 \Gamma \overrightarrow{D} S^2  \pm S^2 \overleftarrow{D} \Gamma S^1) \psi \;,
\label{eq:der_op2}
\ee
where the ``$\pm$'' symmetrization ensures a good $C$-parity quantum number. Following
Eq.~(\ref{eq:der_op2}), some interpolators (with $S^1=S^2$) are identical to zero
after partial integration. The operator Eq.~(\ref{eq:der_op1}) is in general
non-vanishing even if $S^1=S^2$, since $[D,S]\neq0$. 
This commutator can be seen as introducing additional pieces of paths in the combined smearing operator, 
which means changed weights of the existing paths and a few new paths.
Numerically, we find that
the corresponding correlators are of the same magnitude as others and yield
consistent signals. Hence, this asymmetric definition 
enlarges effectively the basis of operators to some extent.
In particular some exotic channels can be accessed this way already 
with fewer derivatives.

In Appendix \ref{interpolators} we list all meson interpolators used in our
study, ordered according to their spin and parity.  The tables differ from those 
in \cite{Engel:2010my} since we here account for the approximate symmetry under 
$C$-parity of strange mesons and  construct the interpolators accordingly.
Monitoring the eigenvectors in the variational method allows for
insights in approximate $C$-parities of various strange meson states, and
furthermore in the breaking of $C$-parity of strange mesons 
when approaching the physical pion mass.

\section{Isovector light mesons}\label{lightmesons}

\begin{figure}
\includegraphics[width=\columnwidth,clip]{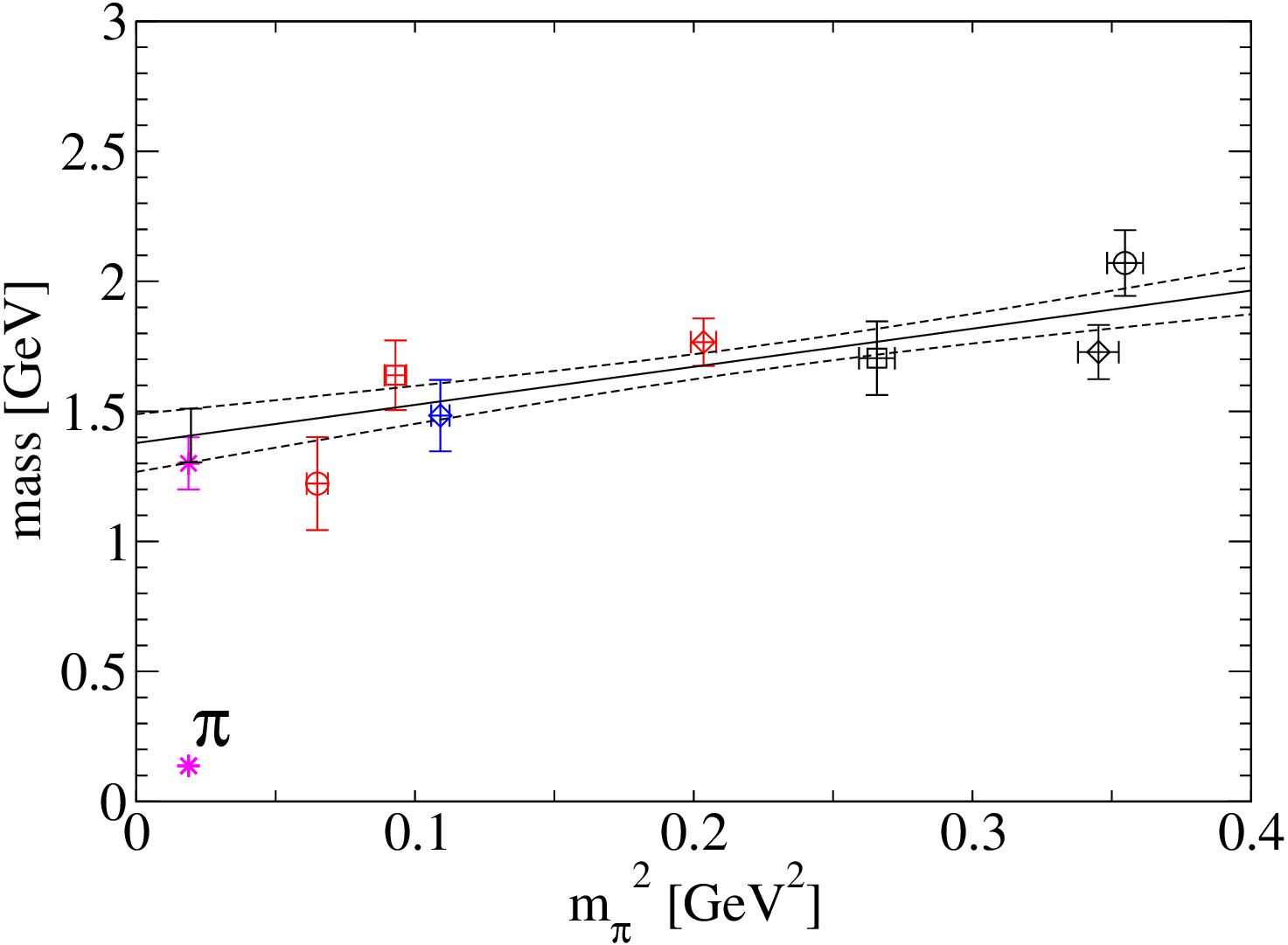}\\
\noindent\includegraphics[width=\columnwidth,clip]{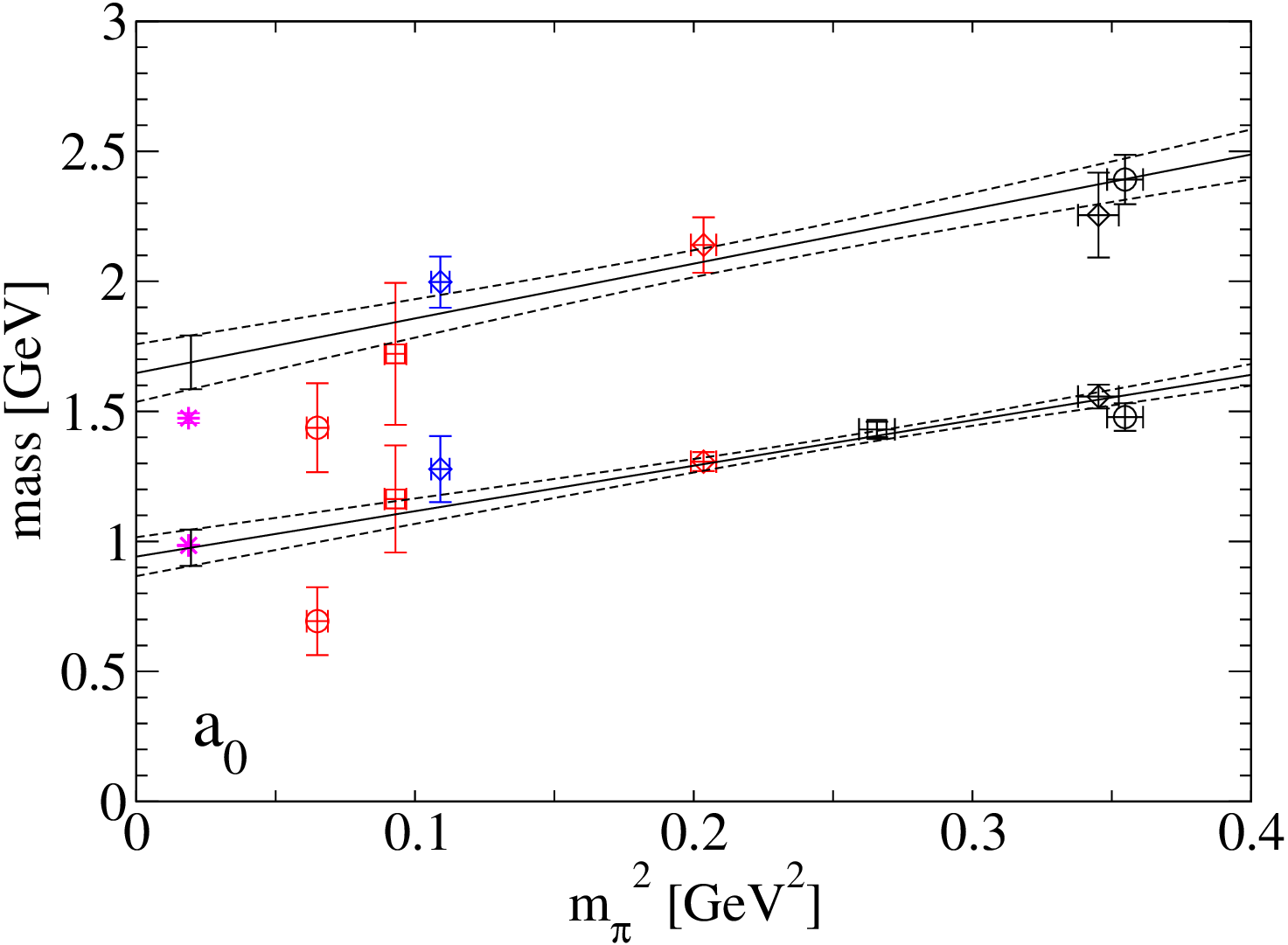}
\caption{$0^{-+}$ (top): only the first excitation is shown, the ground state
pion mass squared defines the abscissa. $0^{++}$
(bottom):  observed ground state and first excitation.}
\label{mesons:0}
\end{figure}

The energy levels are obtained from exponential fits to the eigenvalues in a
range of $t$-values where the eigenvalues and eigenvectors are compatible with
plateau behavior. Typically that plateau extends from $t=2$ or 3 up to $t=6$ 
to 12. In some cases the eigenvalues are close to each other and their order
changes  from one time\-slice to another  and also changes randomly over the set
of configurations.  This complicates the exponential fits to the eigenvalues and
the  automatic attribution of the eigenvectors to  physical eigenstates.  In
such situations we use scalar products of eigenvectors at a given timeslice with
the eigenvectors at the preceding timeslice to sort the eigenvalues according to
their corresponding physical states. 
This procedure becomes more reliable towards finer lattice spacings.
For subsets of configurations (in the
jackknife analysis) the eigenvectors are  contracted with the average of the
vectors at the same timeslice.

All masses are extrapolated towards the physical point as a function of the
pion ground state mass squared. In the plots we also show the corresponding
one $\sigma$ error band (dashed curves). The number of energy levels shown is always less than the number of
interpolators chosen for the diagonalization.
The $\chi^2$ per degree of freedom for the chiral fits of all energy levels are collected in the Tables \ref{tab:chi2lightmesons}, and \ref{tab:chi2strangemesons}.
\ref{tab:chi2isoscalarmesons}

\subsection{Scalars}

\myparagraph{0^{-{+}} (\pi)} For the first excitation in the pion channel  (see
Fig. \ref{mesons:0}), the set of operators (1,2,17) is used in all ensembles.
The corresponding effective mass plateaus are rather short, increasing the
uncertainty of the extracted mass. Due to the finiteness of the lattice, the
back-running pion limits the possible fit range for the first excitation
\cite{Gattringer:2008be,Gattringer:2008vj,Prelovsek:2010kg}, in particular at
small pion masses.  Nevertheless, masses can be extracted and the chiral
extrapolation hits the experimental $\pi(1300)$ within $1\sigma$.

\begin{figure}[t]
\noindent\includegraphics[width=\columnwidth,clip]{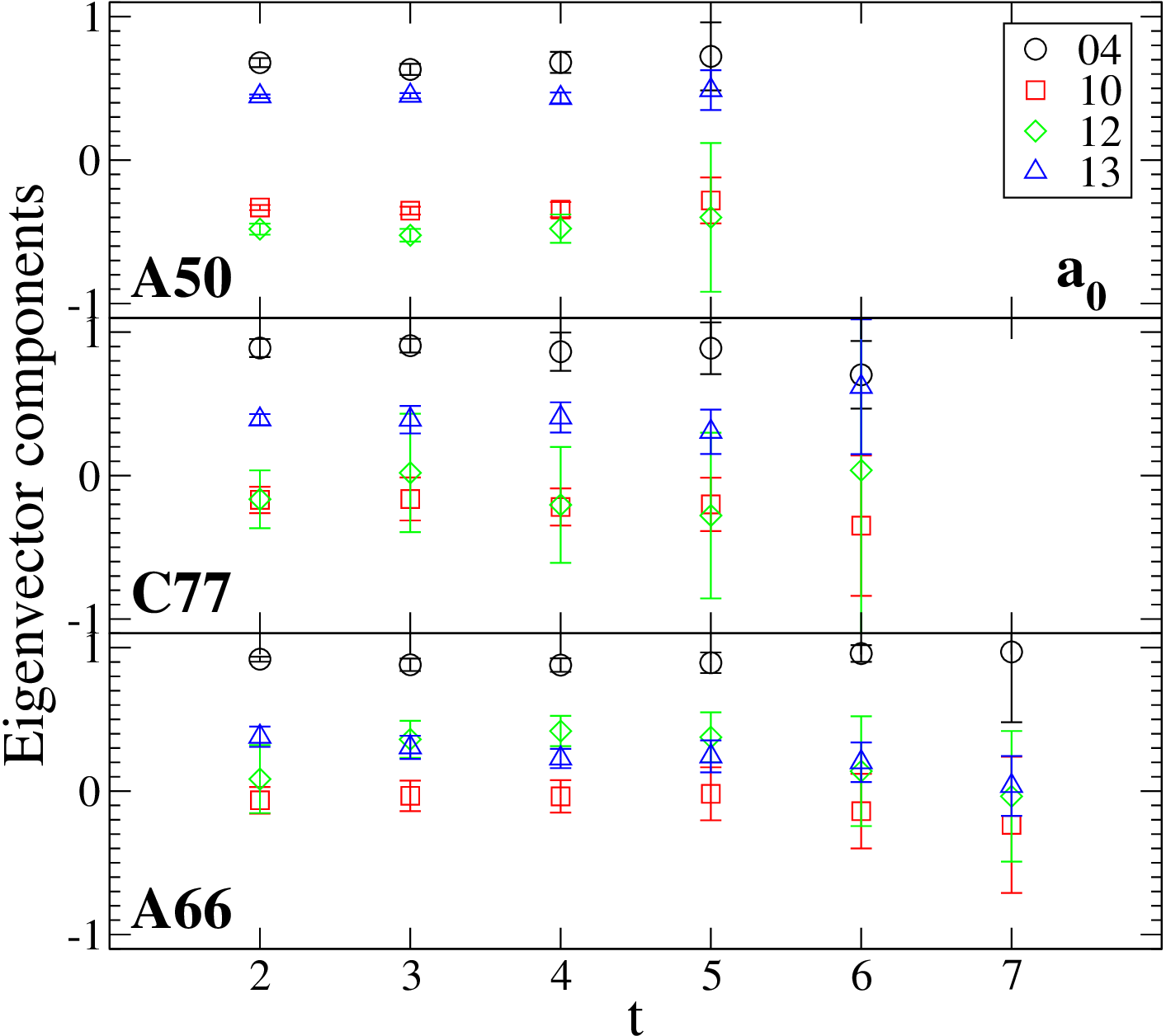}
\caption{Eigenvector components of the ground state of the light scalar meson channel ($a_0$) 
of ensembles A50, C77 and A66 (top to bottom).
Interpolator (4) (only gaussian sources) dominates, while 
contributions of the other interpolators (one or two derivatives) is found to be particularly relevant at heavy pions.
Nevertheless, the eigenvectors are very similar over the whole range of pion masses (600 to 250 MeV) and only evolve smoothly.}
\label{evecs_a0}
\end{figure}

\begin{figure}[t]
\noindent\includegraphics[width=\columnwidth,clip]{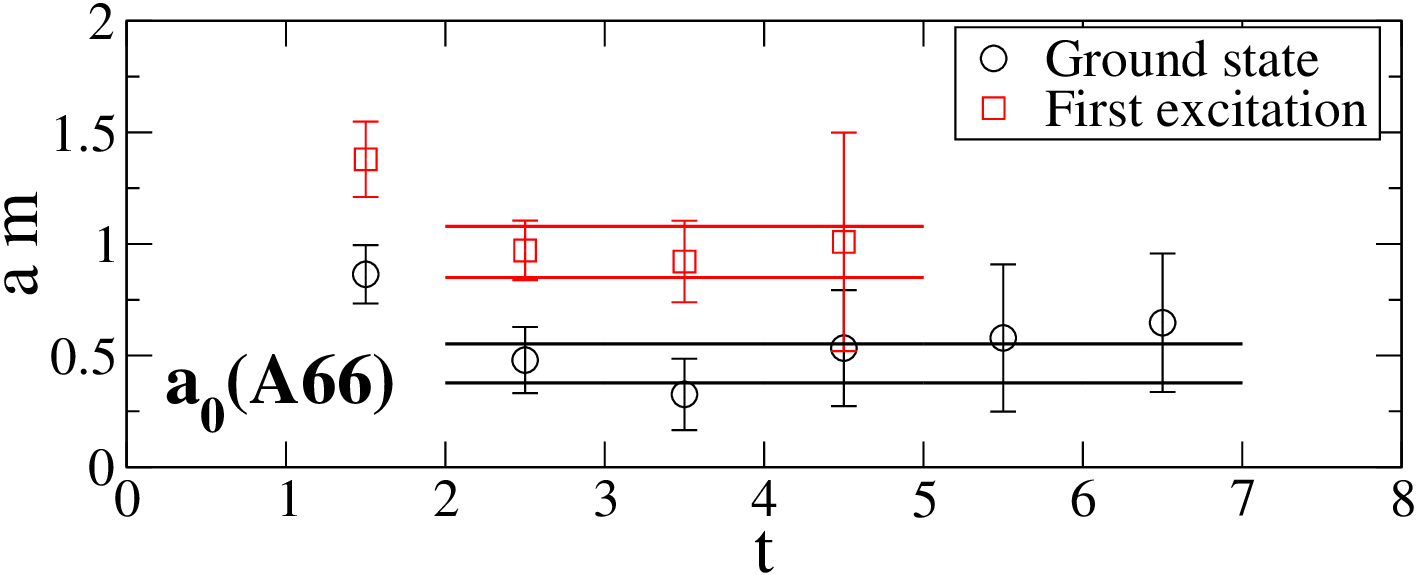}
\caption{Effective mass plateaus for the  light scalar meson channel ($a_0$) of ensemble A66, 
ground state and first excitation.}
\label{effmass_a0}
\end{figure}

\myparagraph{0^{+{+}} (a_0)}
In \cite{Engel:2010my} three (A50, B70 and C77) of the seven ensembles  have
been analyzed, with less statistics than in the present work. Partially
quenched data was used to argue that the signal in the $0^{++}$ channel
probably has significant contributions from the S-wave scattering state
$\pi\eta_2$. In the present work we analyze only fully dynamical data  (except
for the strange sector). Our results are now compatible with the experimental ground state $a(980)$ within 1~$\sigma$ and with the first excitation
$a(1450)$ within 2~$\sigma$ (see Fig. \ref{mesons:0}). However, the channel still poses some difficulties. The
plateau is rather short and there remains some ambiguity in choosing the fit
range, leading to a systematic error. In addition, the results depend on the
chosen set of interpolators.  We show results from subsets of 
(1,4,10,12,13). In ensemble B60, the excitation signal was not
good enough to be fitted.
The extrapolations of the
ground state levels agree for the different choices of interpolators.

However, in particular the ground state energy level of ensemble A66 deviates
when changing the set of interpolators. The result becomes unexpectedly light, most pronounced in the case of the set (10,12,13), though 
the corresponding effective mass plateaus look stable. Indeed, this point lies below the (theoretical)
$\pi\eta_2$ threshold and could indicate a scattering state signal.
It also could signal a severe finite size effect for this case in A66; this
could be clarified only by increasing the lattice volume. Nevertheless,
except for this point, the results are compatible with the experimental
states.

In Fig.~\ref{evecs_a0} we show the eigenvectors for the ground state for three
ensembles covering the whole range of pion masses presented. They are quite consistent
with each other and not supporting the notion of a change in the physics of the ground 
state over that range. Fig.~\ref{effmass_a0} shows the effective masses of ground
state and first excited energy level for the ensemble with smallest pion mass (A66).

There are studies for the finite size dependence of the lowest 
energy level in this
channel based on unitarized chiral perturbation theory \cite{Doring:2011vk}.
However, at the moment our values are not precise enough to
decide on these grounds on properties of the $a_0$. Also it may be
necessary to include meson-meson interpolators in a more detailed study.
Simulations to address finite-size effects are currently in progress
and the discussion of this ongoing effort is beyond the scope of the
current publication.

\begin{figure*}[t]
\noindent\includegraphics[width=\columnwidth,clip]{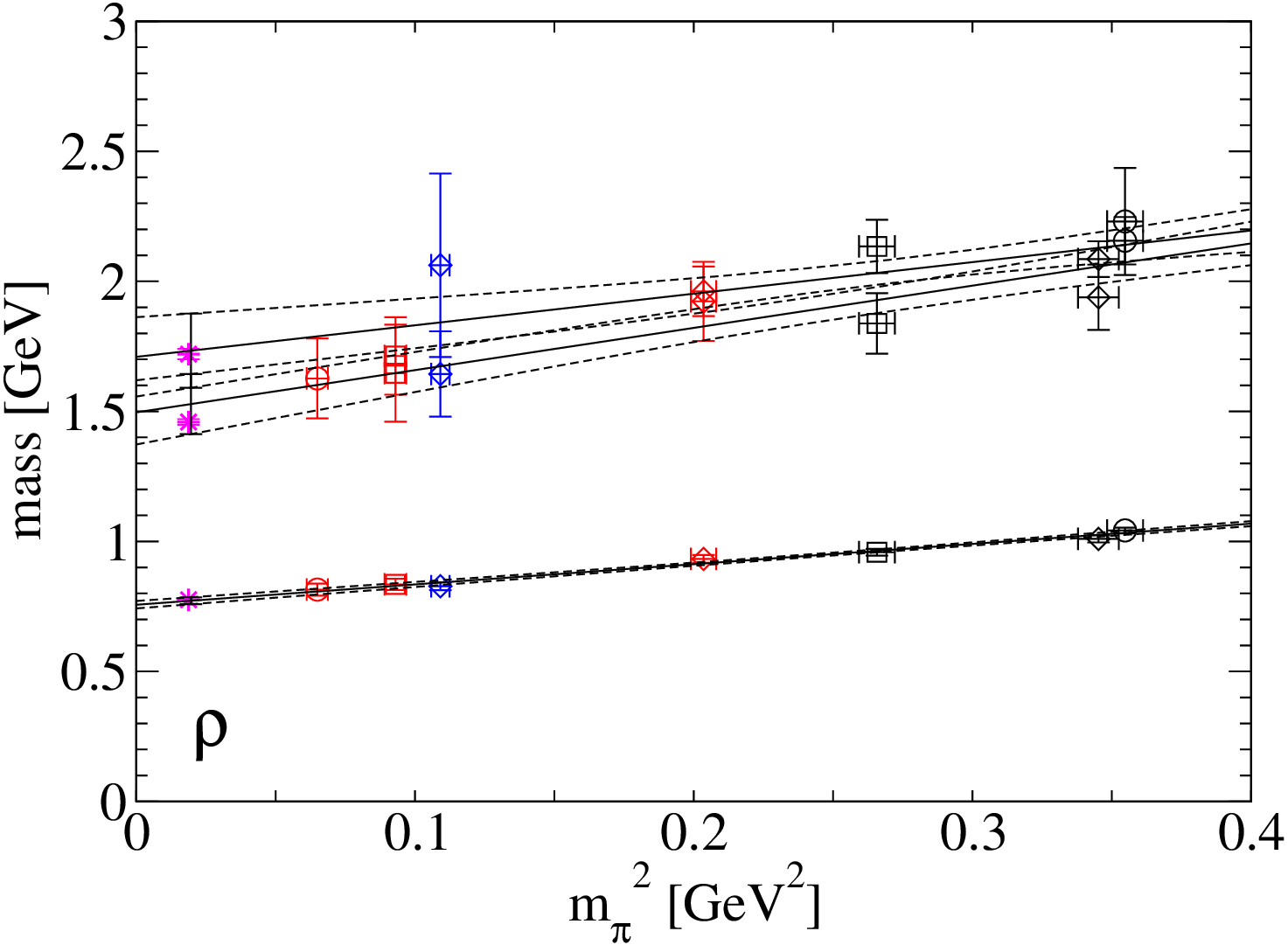}
\noindent\includegraphics[width=\columnwidth,clip]{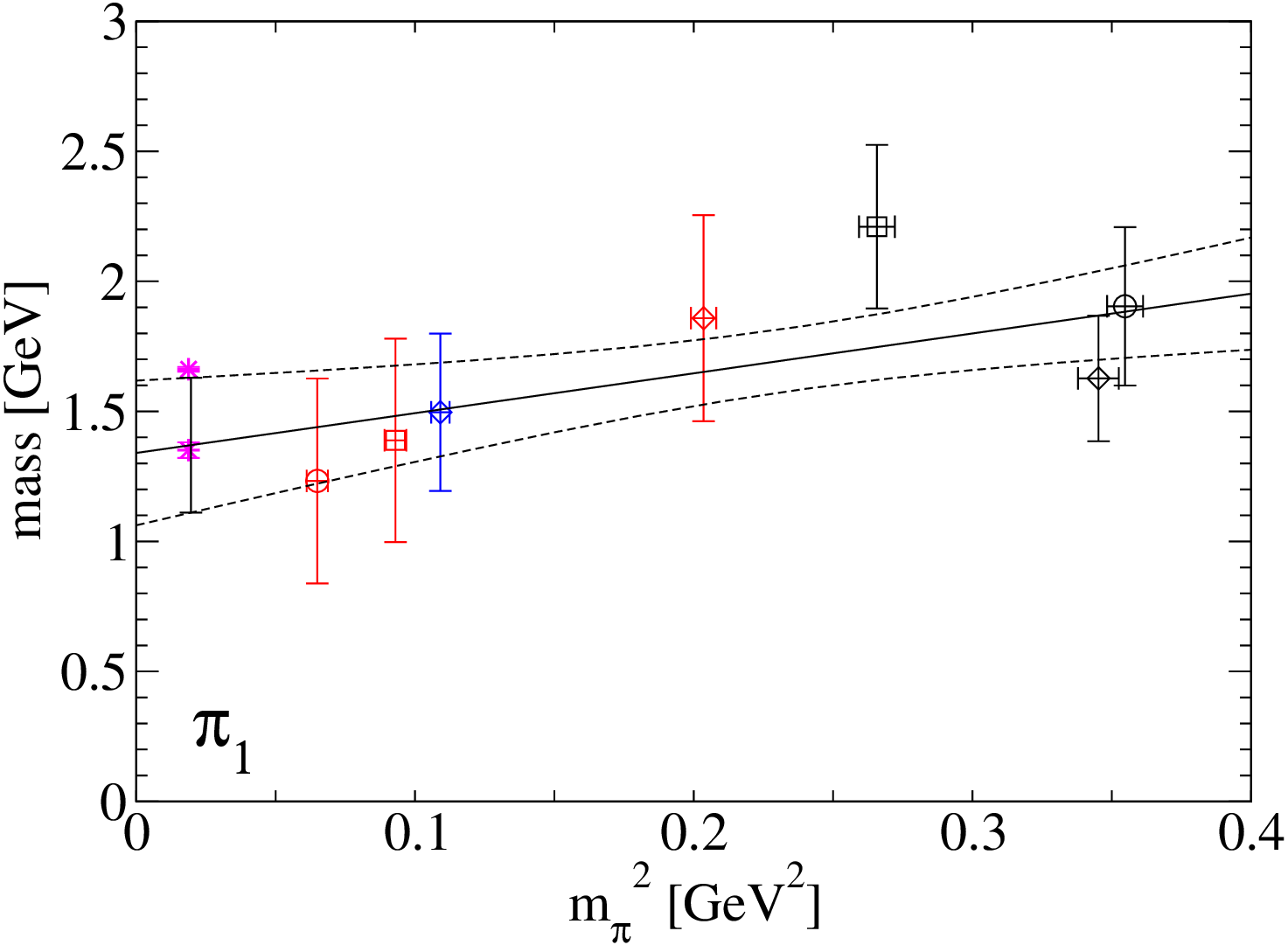}\\
\noindent\includegraphics[width=\columnwidth,clip]{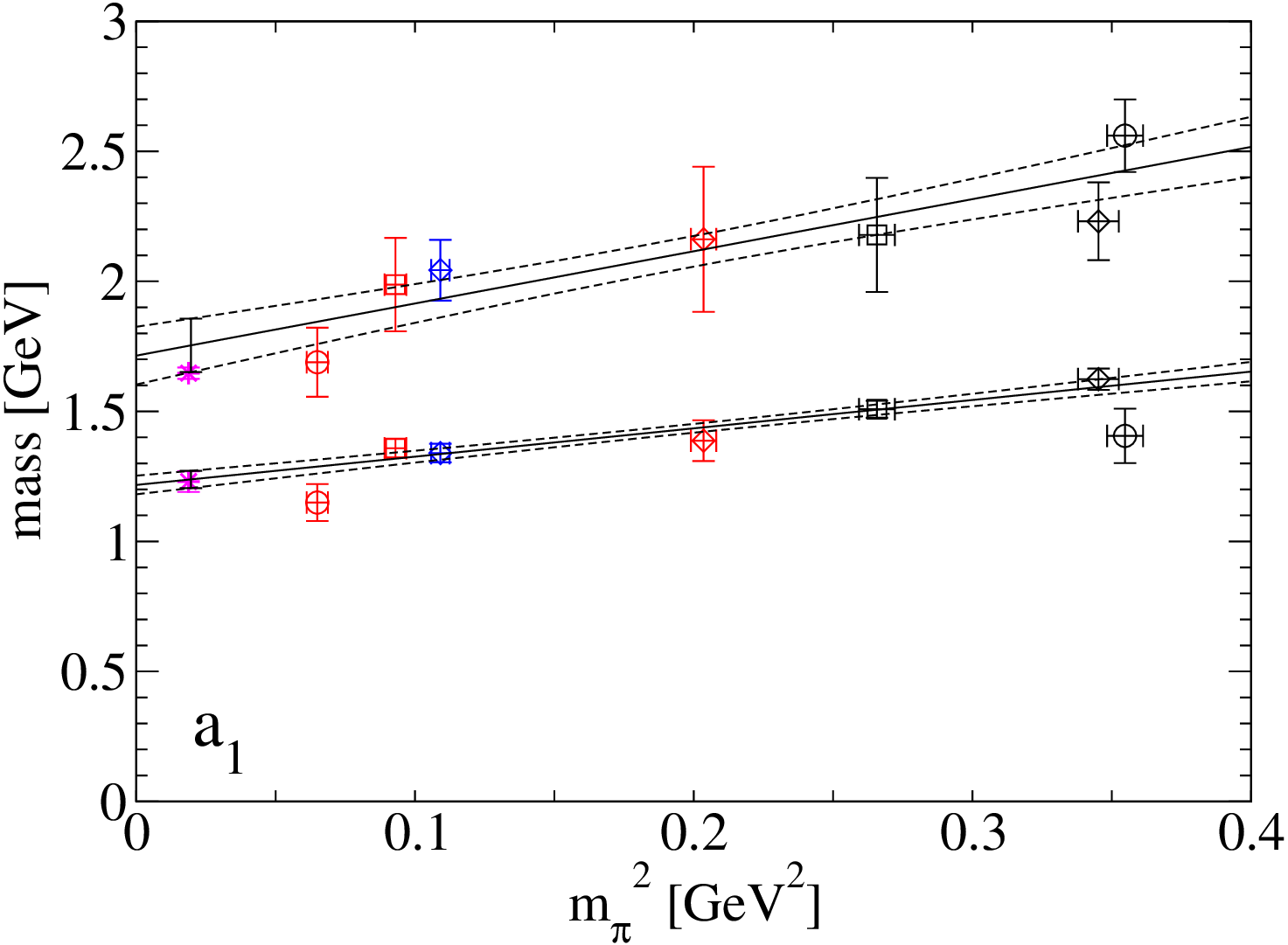}
\noindent\includegraphics[width=\columnwidth,clip]{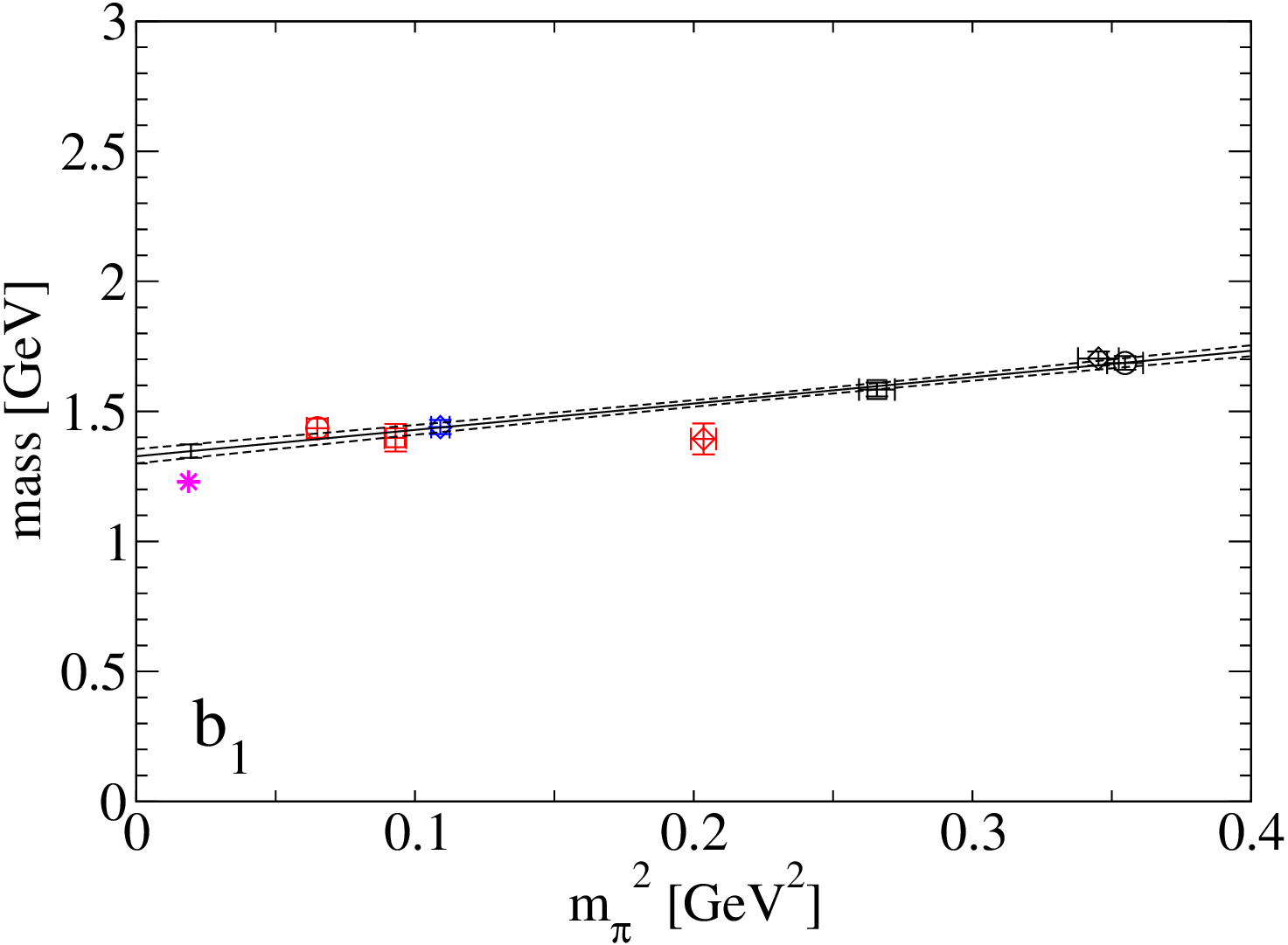}
\caption{(top left)$1^{--}$ ($\rho$); (top right)$1^{-+}$ ($\pi_1$); 
(bottom left)$1^{++}$ $(a_1)$; (bottom right)$1^{+-}$ $(b_1)$; for discussion refer to the text.
}
\label{mesons:1}
\end{figure*}

\begin{figure}
\noindent\includegraphics[width=\columnwidth,clip]{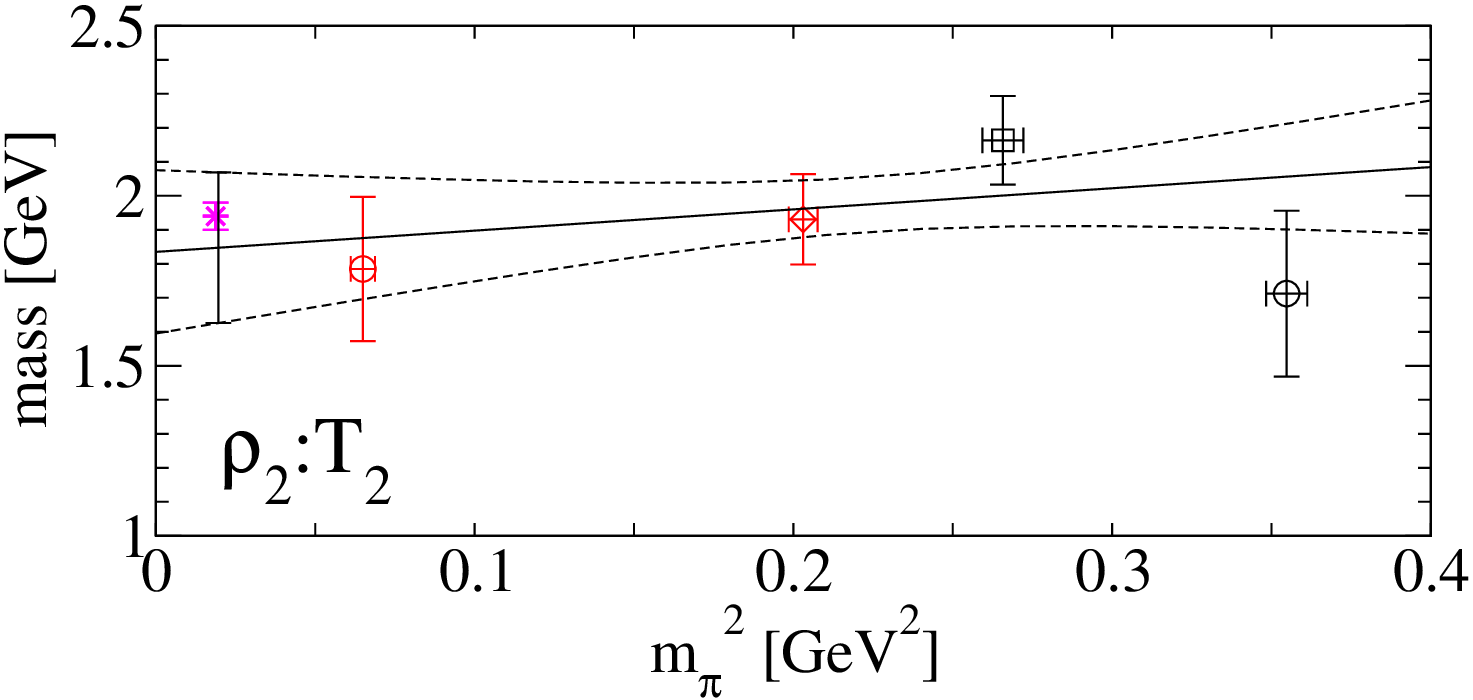}\\
\noindent\includegraphics[width=\columnwidth,clip]{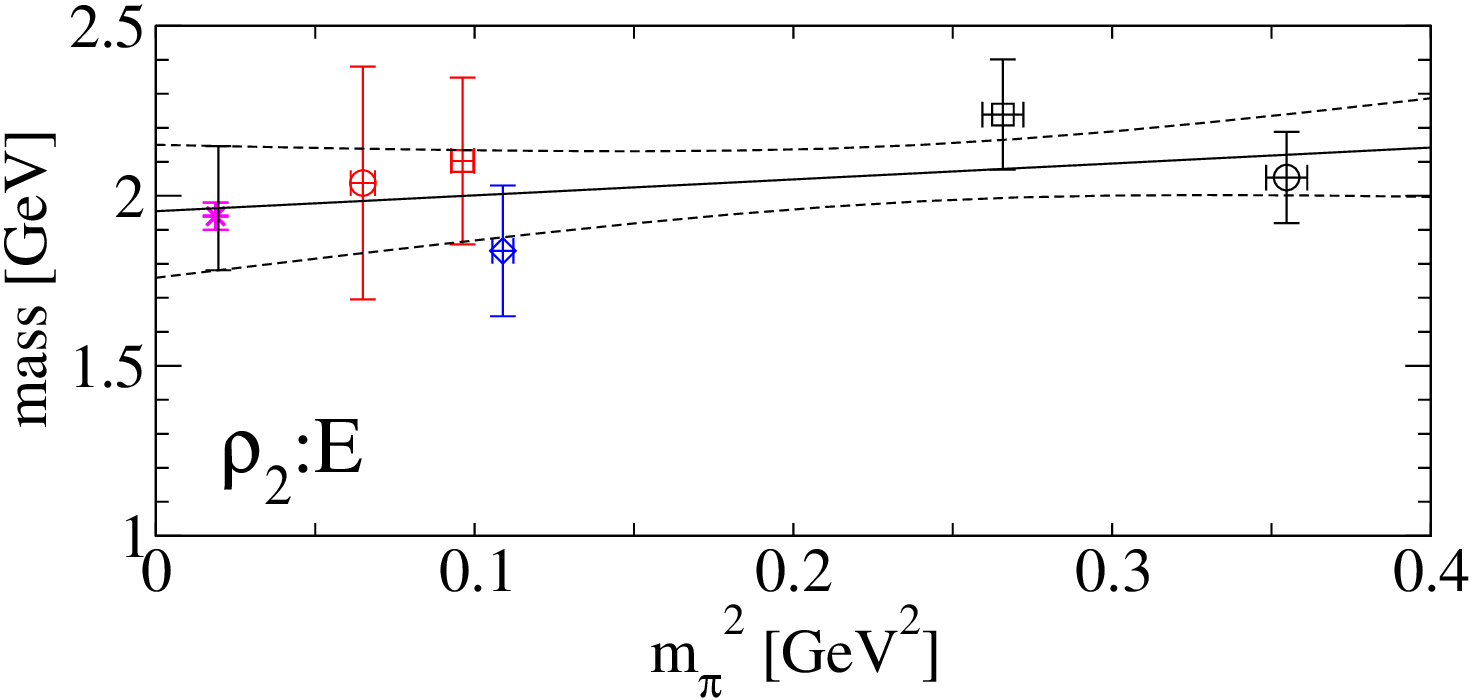}\\
\noindent\includegraphics[width=\columnwidth,clip]{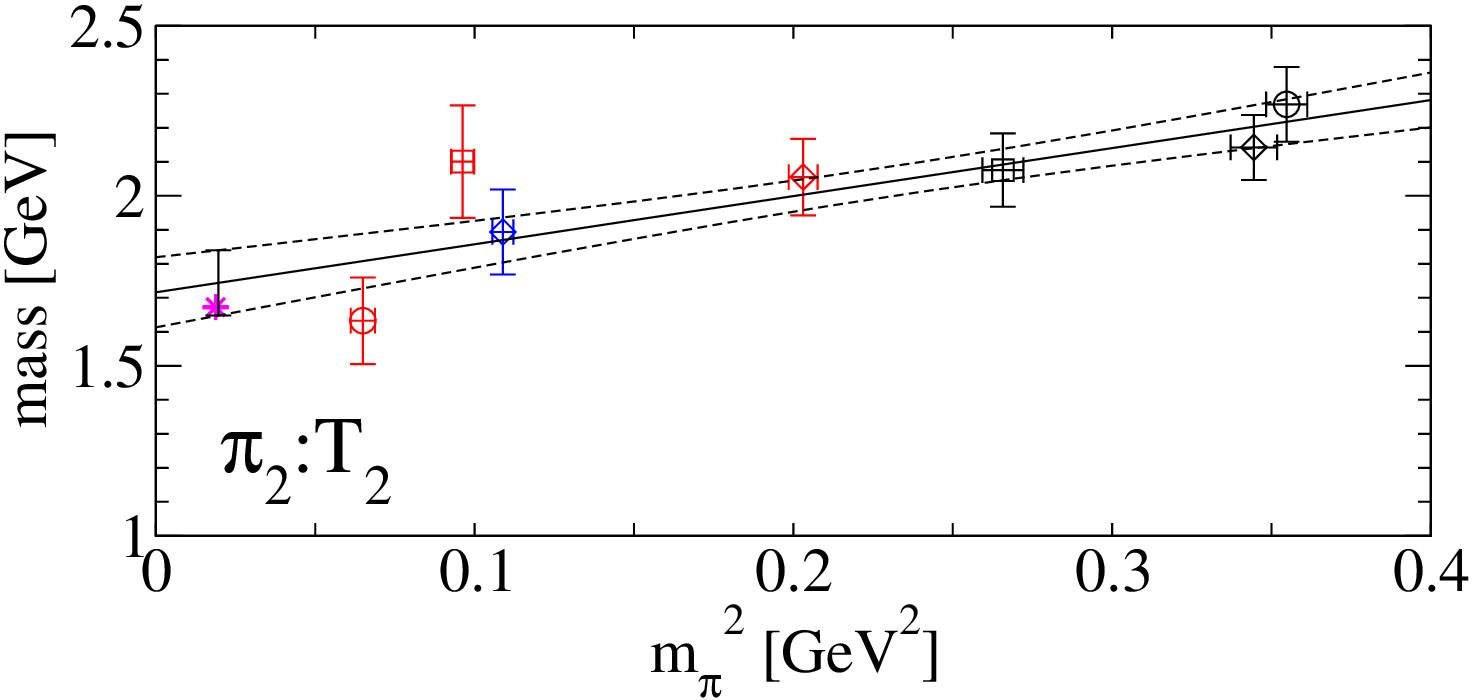}
\caption{(top and middle) $2^{--}$ ($\rho_2$) in both representations; 
(bottom) $2^{-+}$ ($\pi_2$) in representation $T_2$.}
\label{mesons:2}
\end{figure}
\subsection{Vectors}\label{light-vectors}

\myparagraph{1^{-{-}} (\rho)}
The $\rho(770)$ comes out nicely as usual (see Fig.~\ref{mesons:1}). 
The first and second excitation are extracted using the set (1,8,12,17,22),
where the second excitation is not stable in A66. 
These excitations are very close to one another, making the chiral
extrapolations less reliable. The pattern of energy levels would 
allow a crossover of eigenstates
but the eigenvectors do not confirm this. Therefore, we extrapolate
the results to the physical point according to the na\"ively assumed level
ordering, neglecting a possible crossover. The results are compatible with the
experimental $\rho(1450)$ and $\rho(1570$ or $1700)$ within error bars (for a
discussion on the latter excitation see \cite{Nakamura:2010zzi}). 

We find no obvious indication for a coupled $\pi\pi$ P-wave channel. As discussed
earlier \cite{Engel:2010my,Bulava:2010yg} this may be due to weak coupling. By including two pion interpolators one can derive a scattering phase shift from the
modification of the observed energy levels close to the resonance (see, e.g. \cite{Lang:2011mn}). Such a study needs inclusions of disconnected graphs, which
are not accessible to us: The necessary propagator calculation is numerically
too costly for CI fermions.

\myparagraph{1^{-{+}} (\pi_1)}
The quantum numbers $1^{-{+}}$ cannot be obtained with isotropic quark sources
only. Thus, this channel is not accessible by simple quark models, and it is
commonly referred to as exotic. Due to the weak signal, the set of operators
has to be optimized in each ensemble separately, taking one or two 
interpolators of (9,11,14,16,21,24). This way a mass value can be extracted
only with comparatively large statistical uncertainty. 
The chiral extrapolation hits the experimental $\pi_1(1400)$, but is also compatible with the $\pi_1(1600)$  (see Fig.~\ref{mesons:1}).
In some of the ensembles we get the best signal using interpolators  which are nonzero
only due to the definition in Eq.~\ref{eq:der_op1} and discussed there. This
may be related to the ``exotic'' property of this channel.

\myparagraph{1^{+{+}} (a_1)}
The signal in the pseudovector meson channels is usually bad compared to the
pion and the $\rho$ channels. Nevertheless, the ground state and a first
excitation can be identified. The ground state is extracted using the single
interpolator (1). For the first excitation the set has to be optimized in each
ensemble separately, taking subsets of three interpolators out of
(1,2,4,13,15,17). Some of the plateaus tend to shift towards smaller masses
 at large time separations. 
However, as far as possible, long fit ranges are chosen.
The chiral extrapolations hit the experimental
$a_1(1260)$ and the $a_1(1640)$ within error bars  (see Fig.~\ref{mesons:1}).

\myparagraph{1^{+{-}} (b_1)}
In the $1^{+{-}}$ channel, the ground state plateau is more stable than in its
positive $C$-parity partner channel ($a_1$). Using the single interpolator
(6), a mass with comparatively small error bar is obtained. The chiral
extrapolation comes out  too high, missing the experimental $b_1(1235)$ by
more than $2\sigma$  (see Fig.~\ref{mesons:1}).

\subsection{Tensors}

The continuum representation for spin 2 contributes to
the irreducible representations $T_2$ and $E$ on the lattice. These interpolators are
orthogonal, thus masses can be extracted in each of them
separately. In the continuum limit, the results should agree, however, at
finite lattice spacings they can show different discretization effects. We
extract the energy levels separately  and compare the corresponding chiral
extrapolations.

\myparagraph{2^{--} (\rho_2)}
In many of the spin 2 channels the signal is weak and fits can be performed
only for some of the seven ensembles. In particular this is the case in the
$2^{-{-}}$ channel (see Fig. \ref{mesons:2}, top and middle). We use the single interpolator
(2) in $T_2$ and also (2) in $E$. The effective masses are noisy, 
the fitted plateaus are rather short, with only 2 d.o.f.~in the fits.
Nevertheless, the chiral extrapolations of the $T_2$ and
$E$ ground state masses agree with each other and also with the experimental $\rho_2(1940)$ mass.
Hence, our results are compatible with this state, which is omitted from the
summary table of \cite{Nakamura:2010zzi}.

\myparagraph{2^{-+} (\pi_2)}
In the $2^{-{+}}$ channel  (Fig. \ref{mesons:2}, bottom), interpolator (6) is
applied in $T_2$. The extrapolation to the physical point
is compatible with the experimental
$\pi_2(1670)$ (within 1 resp.~1.5$\sigma$). 
The signal for representation $E$ (not shown) is too weak to
be reliable.  

\myparagraph{2^{+-}}
We studied this channel for completeness but the signals were inconclusive and
did not allow to extract an energy level.

\begin{figure}
\noindent\includegraphics[width=\columnwidth,clip]{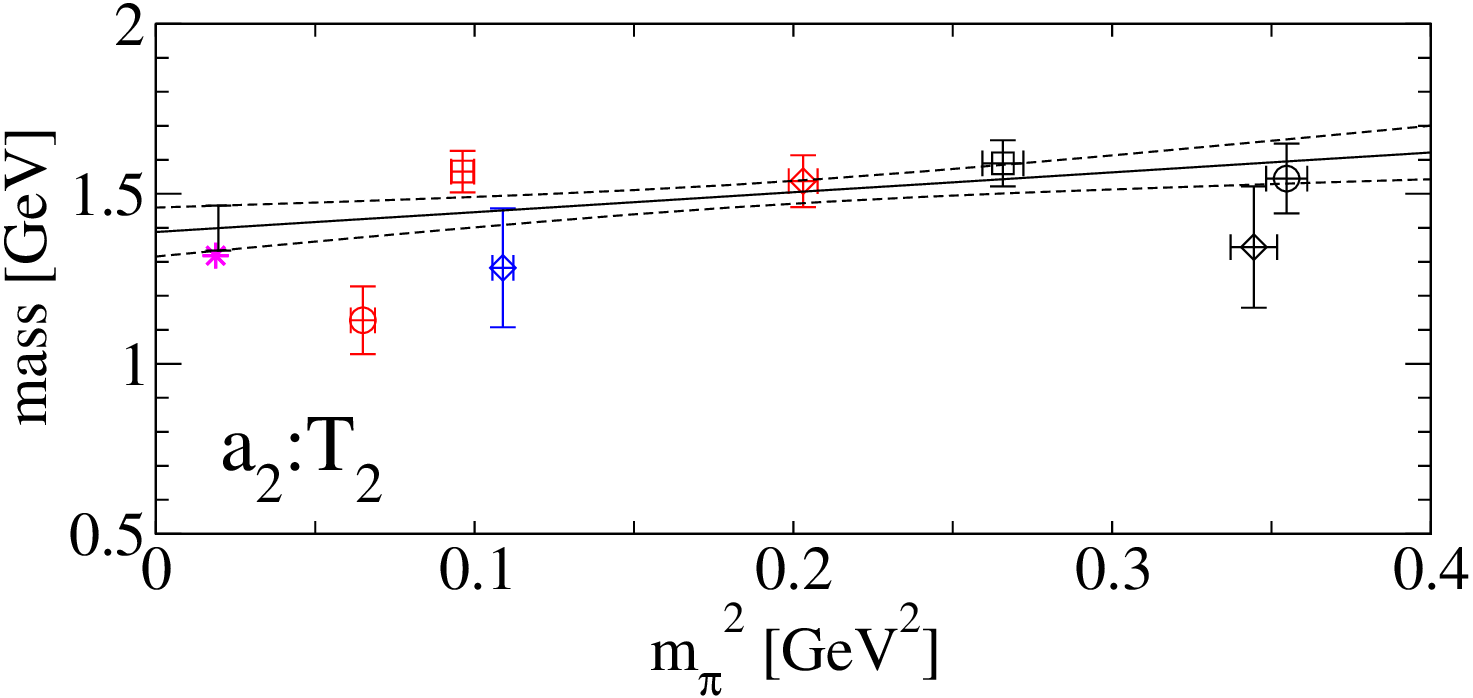}\\
\noindent\includegraphics[width=\columnwidth,clip]{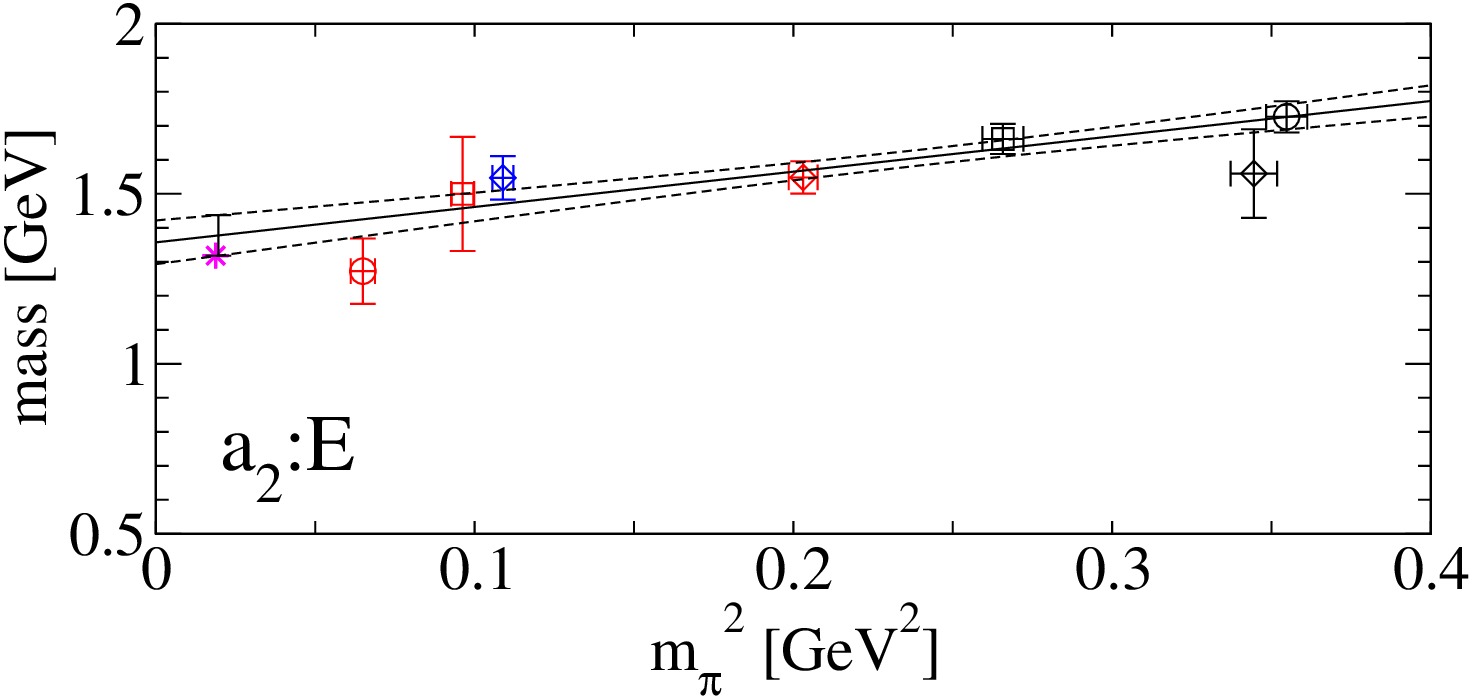}
\caption{$2^{++}$ ($a_2$) in both representations.}
\label{mesons:2++}
\end{figure}

\myparagraph{2^{++} (a_2)}
In the $2^{+{+}}$ channel (Fig. \ref{mesons:2++}), we use interpolator (2) in
$T_2$ and (2) (respectively (6) for A66) in $E$. Some of the plateaus are
unexpectedly light, however, that might be statistical
fluctuation. The chiral extrapolations of the $T_2$ and $E$ ground state masses agree and both match
the experimental $a_2(1320)$ mass within error bars.
The $\chi^2$/d.o.f.~of the chiral fit of $T_2$ is larger than three (see Tab.~\ref{tab:chi2lightmesons}), where the major contribution stems from ensemble A66.
Finite volume effects could be responsible for the significant deviation of this particular value.

\section{Mesons with strange valence quarks}\label{strangemesons}

In 2-flavor simulations, strange hadrons can be studied by including the
strange quark just as a valence quark. The corresponding quantum field theory
is not well defined, the probability distribution of physical observables is not 
anymore strictly non-negative. Nevertheless, since the strange quark is heavy
compared to the light, dynamical quarks, observables can be measured and
regarded as predictions including systematic errors. We stress that even
though light hadrons are well defined in 2-flavor simulations, they also show
the systematic error of neglecting strange sea quarks when the results are
compared to experiment.  From this point of view, the predictive power of
strange valence hadrons is not significantly below the one of light
hadrons in 2-flavor simulations.
The strange quark mass parameter is set in each ensemble such that the
$\Omega(1672)$ is reproduced (always assuming that $r_{0,exp}=0.48\,$fm)
(see Sec.~\ref{subsec:scale_strange}).

\begin{figure}
\noindent\includegraphics[width=\columnwidth,clip]{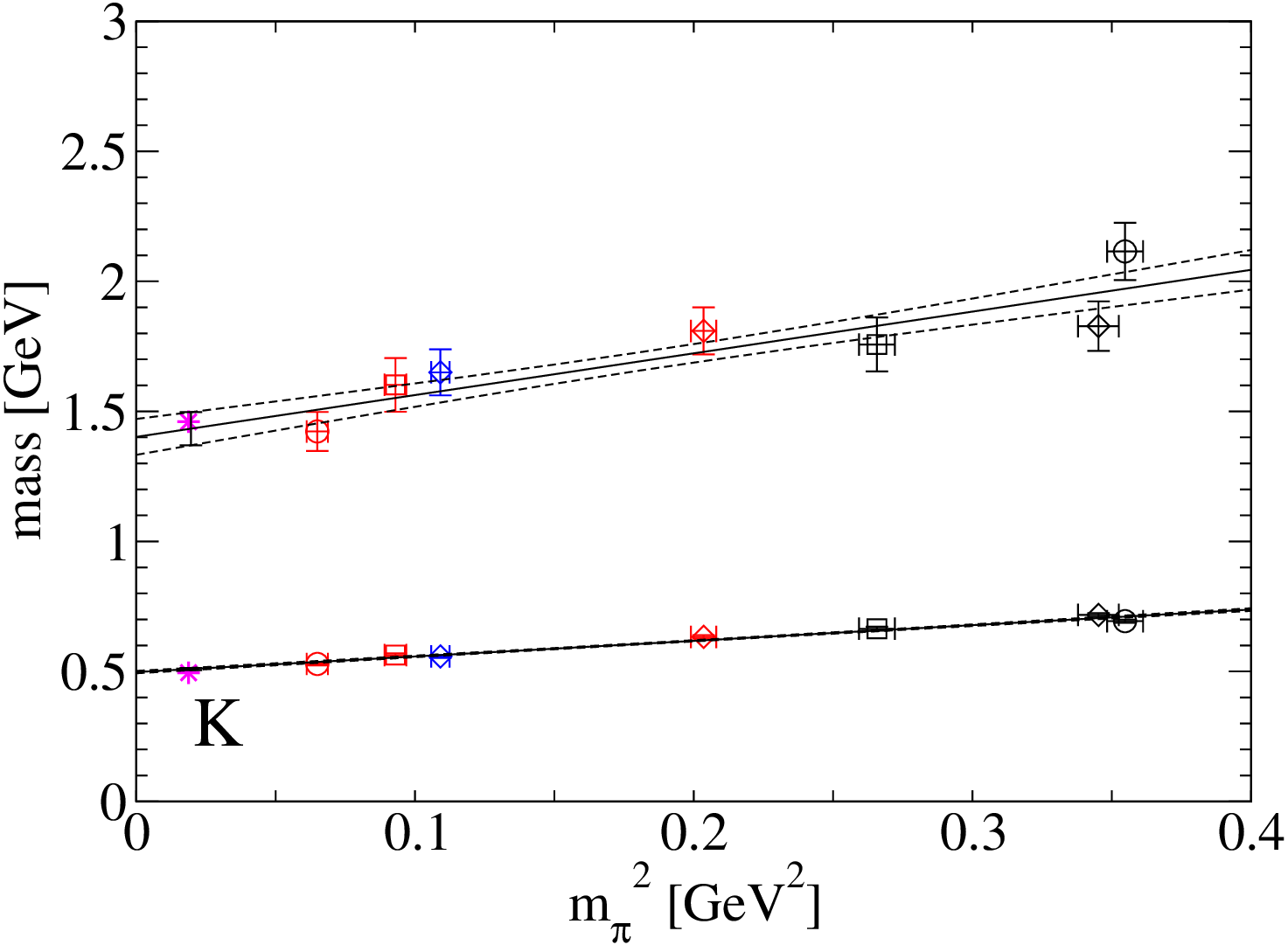}\\
\noindent\includegraphics[width=\columnwidth,clip]{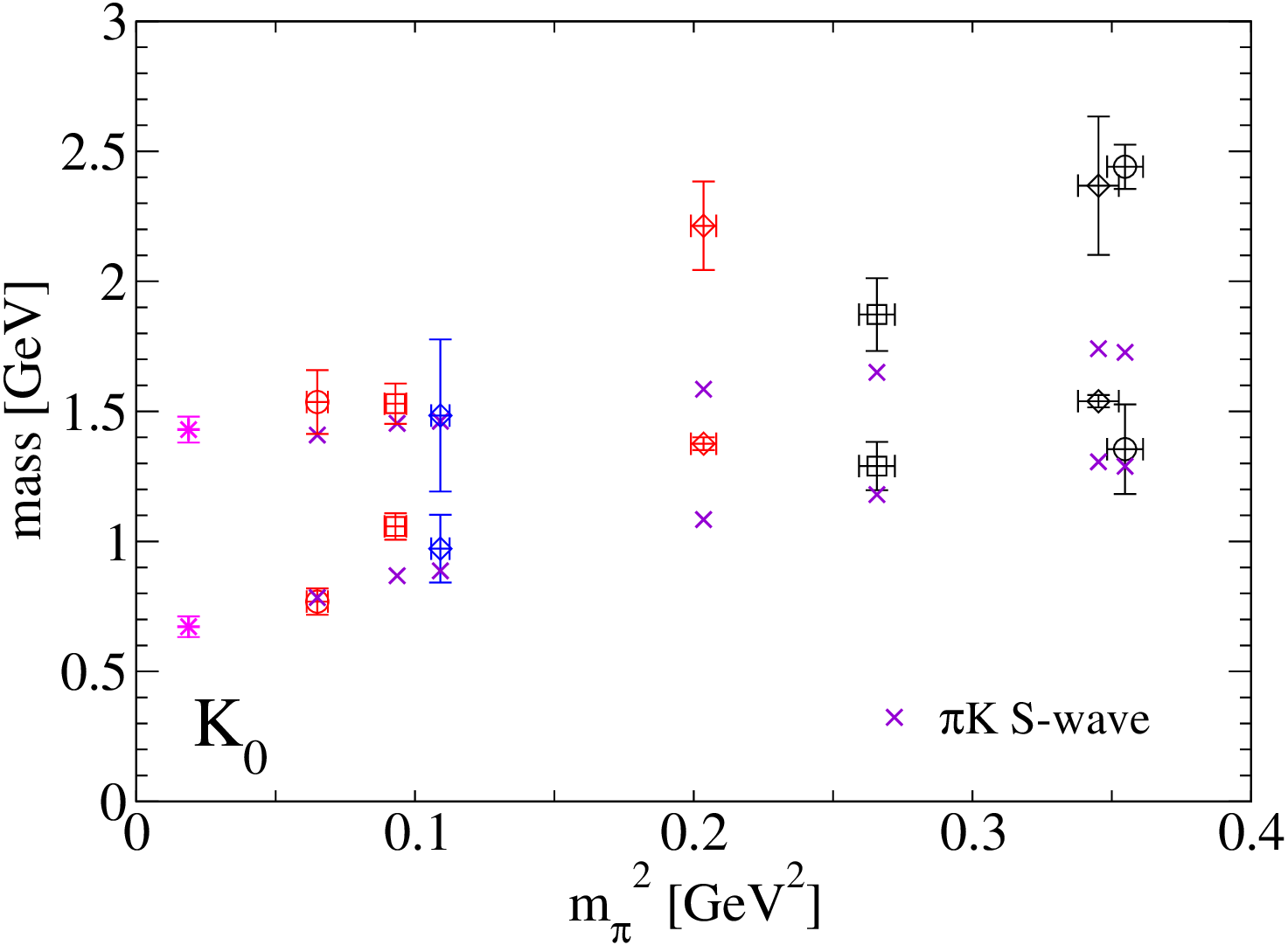}	
\caption{(top) $(I=\frac{1}{2})\;\;0^-$ ($K$). (bottom) $(I=\frac{1}{2})\;\;0^+$ ($K_0$):
The S-wave scattering state $\pi K$ for zero and minimum non-zero relative momentum is indicated for all ensembles using crosses.
The chiral fits are omitted for clarity.}
\label{strange-meson:0}
\end{figure}

In contrast to isovector light mesons, $C$-parity is no good quantum number
for $I=\frac{1}{2}$ strange mesons due to the non-degeneracy of the light and
strange quark mass. At unphysically large pion masses, however, $C$-parity is
approximately restored. Our interpolators (see Appendix \ref{interpolators})
are constructed such that 
$C$-parity is a good quantum number in the limit of degenerate quark masses.
Therefore, by monitoring the eigenvectors of the variational method, we can
learn about the $C$-parity content of the states. 

Since excited states are
always more difficult to deal with than ground states, this raises the demands
on the variational method. In some cases it is therefore suggestive to
separate the channels according to $C$-parity. At our largest pion masses,
around 600 MeV, one expects $C$-parity to be almost restored. Approaching the
physical point, $C$-parity is violated stronger and stronger, and 
the corresponding mixing of interpolators is expected to become increasingly
important. To investigate this mixing, we include all possible
interpolators in the correlation matrix, but we also analyze separately the
sectors with given $C$-parity. The advantage of the second approach is a clearer
distinction of the energy levels, where some come in the $[C=+1]$ sector, some
in the $[C=-1]$ sector. In the combined correlation matrix we see both sets, but
due to the increased noise, fewer levels can be reliably determined.
We discuss this point in the subsequent channels.
Our results for the dominant $C$-parity assignments agree qualitatively 
with \cite{Dudek:2010wm}.
Here we also discuss the corresponding 
mixing, which is accessible due to our lighter pion masses.

\subsection{Scalars}

\myparagraph{(I=\frac{1}{2})\;\;0^{-} (K)}
In the strange $(I=\frac{1}{2})\;\;0^{-}$ channel, interpolator (1) is used
for the ground state, which extrapolates close to the experimental kaon 
(see Fig.~\ref{strange-meson:0}).
The $\chi^2$/d.o.f.~of the chiral fit is larger than four (see Tab.~\ref{tab:chi2strangemesons}), which indicates that due to the tiny statistical errors the
systematic errors (e.g. of setting the strange quark mass) become visible. 
For the excited state, we use the set (1,2,8,17), its linear extrapolation
agrees with the experimental $K(1460)$ within error bars. Hence we can confirm this
state (omitted from the summary table of \cite{Nakamura:2010zzi}). In this
channel we use only $0^{-+}$ interpolators, since the signal of the exotic
$0^{--}$ interpolators is too weak, and the corresponding energy levels lie too
high.

\myparagraph{(I=\frac{1}{2})\;\;0^{+} (K_0)}
The strange scalar channel $0^{+}$ is as peculiar as its light 
multiplet partners. The $K_0^*(800)$ (also called $\kappa$) is a very broad resonance (with
a width of more than 80\% of its mass) and is omitted from the summary table of
\cite{Nakamura:2010zzi} due to its unclear nature.

Using interpolator (13) alone (not shown), the chiral
extrapolation almost hits the presumed center of the resonance. To apply the
variational method, we use the set (10,12,13) and include also (1,4) in the basis at 
small pion masses.
We observe that at light pion masses the effective masses tend to decrease at 
large time separations, which may be a signal for contributions of a scattering state. 
Like in most cases, we choose
a large fit range (e.g., 8 timeslices in A66). 
The results are compatible with
the $K_0^*(800)$ and the $K_0^*(1430)$,
but also with the S-wave scattering state $\pi K$ (see Fig.~\ref{strange-meson:0}).
The $\chi^2$/d.o.f.~of the chiral fit of the ground state is larger than eight (see Tab.~\ref{tab:chi2strangemesons}), which is again interpreted as 
indication for systematic errors, probably related to scattering states.
Here we use only $0^{+{+}}$ interpolators, the signal of
the exotic $0^{+{-}}$ interpolators is too weak.

\subsection{Vectors}

\begin{figure}
\noindent\includegraphics[width=\columnwidth,clip]{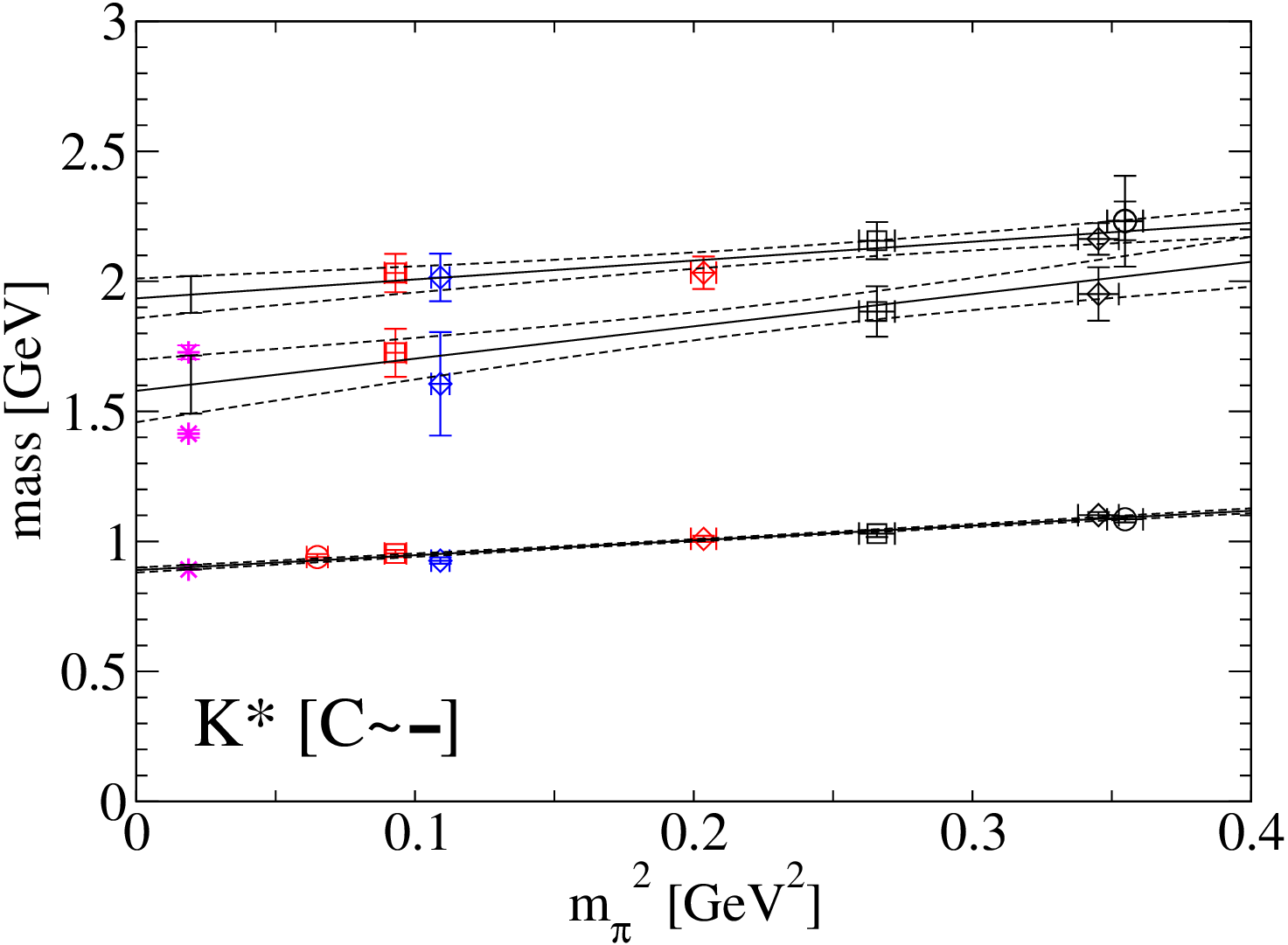}\\
\noindent\includegraphics[width=\columnwidth,clip]{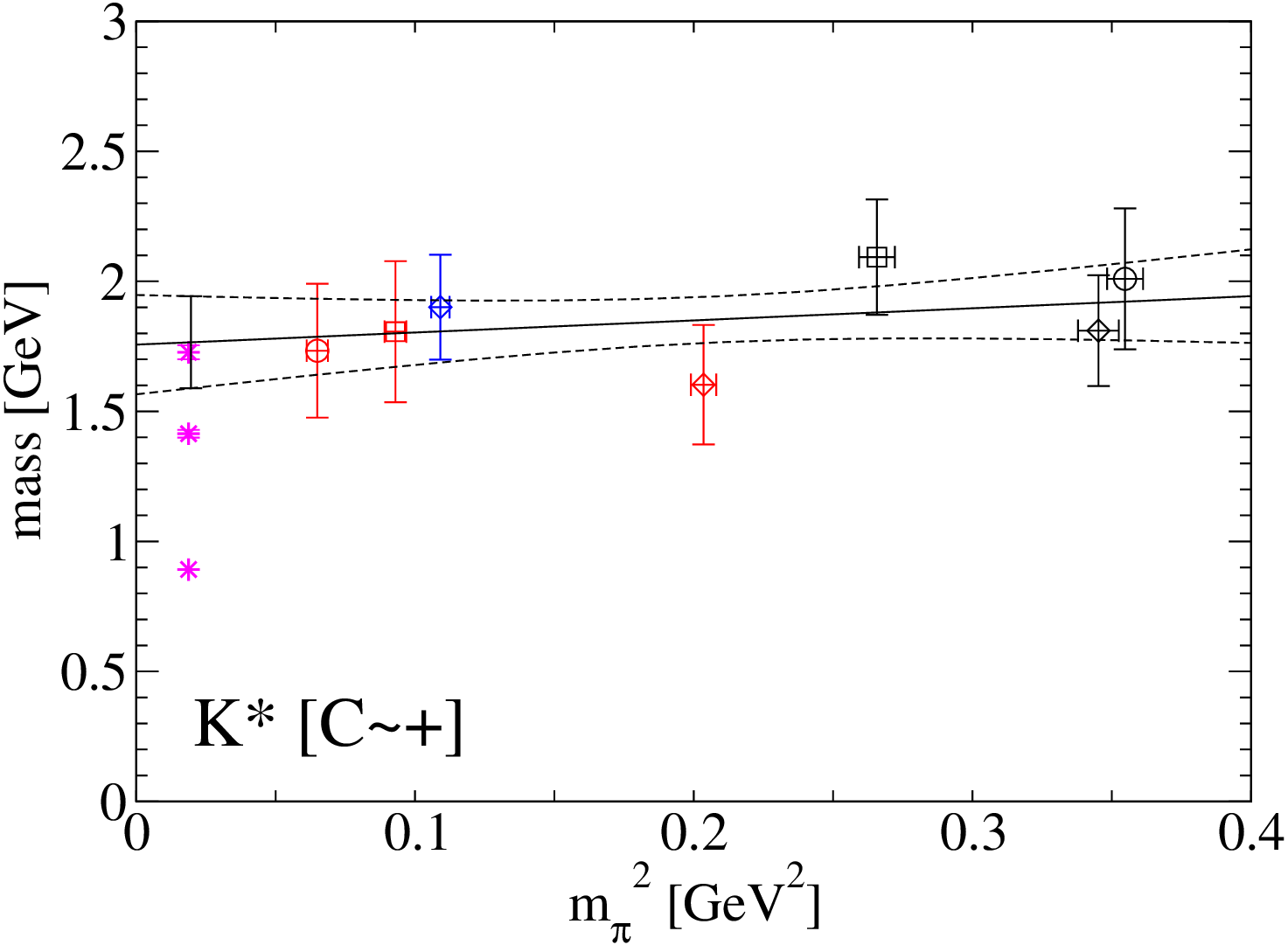}\\
\noindent\includegraphics[width=\columnwidth,clip]{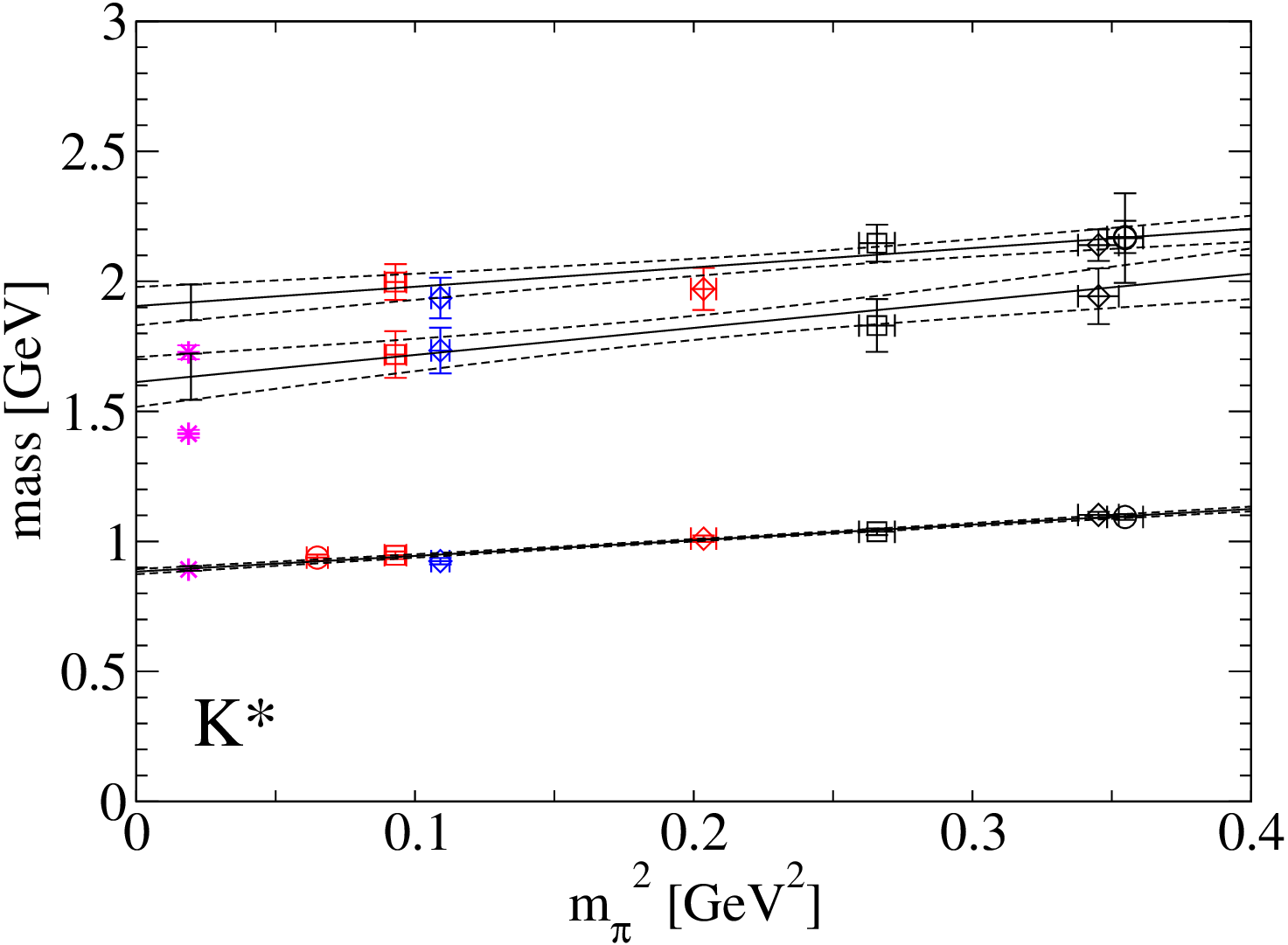}
\caption{$(I=\frac{1}{2})\;\;1^-$ ($K^*$).
Results for interpolators restricted to subsets with $[C\approx -]$  are shown on top and $[C\approx +]$ in the middle.
Note that the ground state is missed in the $[C\approx +]$ subset. 
Results with both types are shown at the bottom.}
\label{strange-mesons:1a}
\end{figure}

\myparagraph{(I=\frac{1}{2})\;\;1^{-} (K^*)}
Considering the strange $J^P$ channels as mixing of $J^{P+}$ and $J^{P-}$, one can
use information from the corresponding light $J^{PC}$ channels  to speculate about 
the dominating $C$-parity in the low-lying states of the strange $J^P$ channel.
Based on that analogy, in
the scalar channels one expects dominance of positive $C$-parity, which is
confirmed by our results. In the vector channels, however, both $C$-parities are
expected to contribute to the measurable low-lying states.
Looking at the experimental states in the corresponding light meson channels
$\rho(770)$, $\pi_1(1300)$, $\rho(1450)$ and $\rho(1570$ or $1700)$, one expects that the
$K^*(892)$ is an (almost) pure $1^{--}$ state, while mixing could become important
for $K^*(1410)$ and $K^*(1680)$.

We first discuss sets of purely negative $C$-parity
interpolators.
Taking interpolators (1,8,12,17,20), we extract a ground state and up to
two excitations. The chiral extrapolation of the ground state hits the
experimental $K^*(892)$ nicely (see Fig.~\ref{strange-mesons:1a}), which is clearly
an (almost) pure $[C\approx-]$ state. The excitations are a bit high compared to the
experimental $K^*(1410)$ and the  $K^*(1680)$.

Considering only $1^{-{+}}$ interpolators, the chiral extrapolation hits the
$K^*(1680)$.   This suggests that mixing is important at least for the $K^*(1680)$.

\begin{figure}
\noindent\includegraphics[width=\columnwidth,clip]{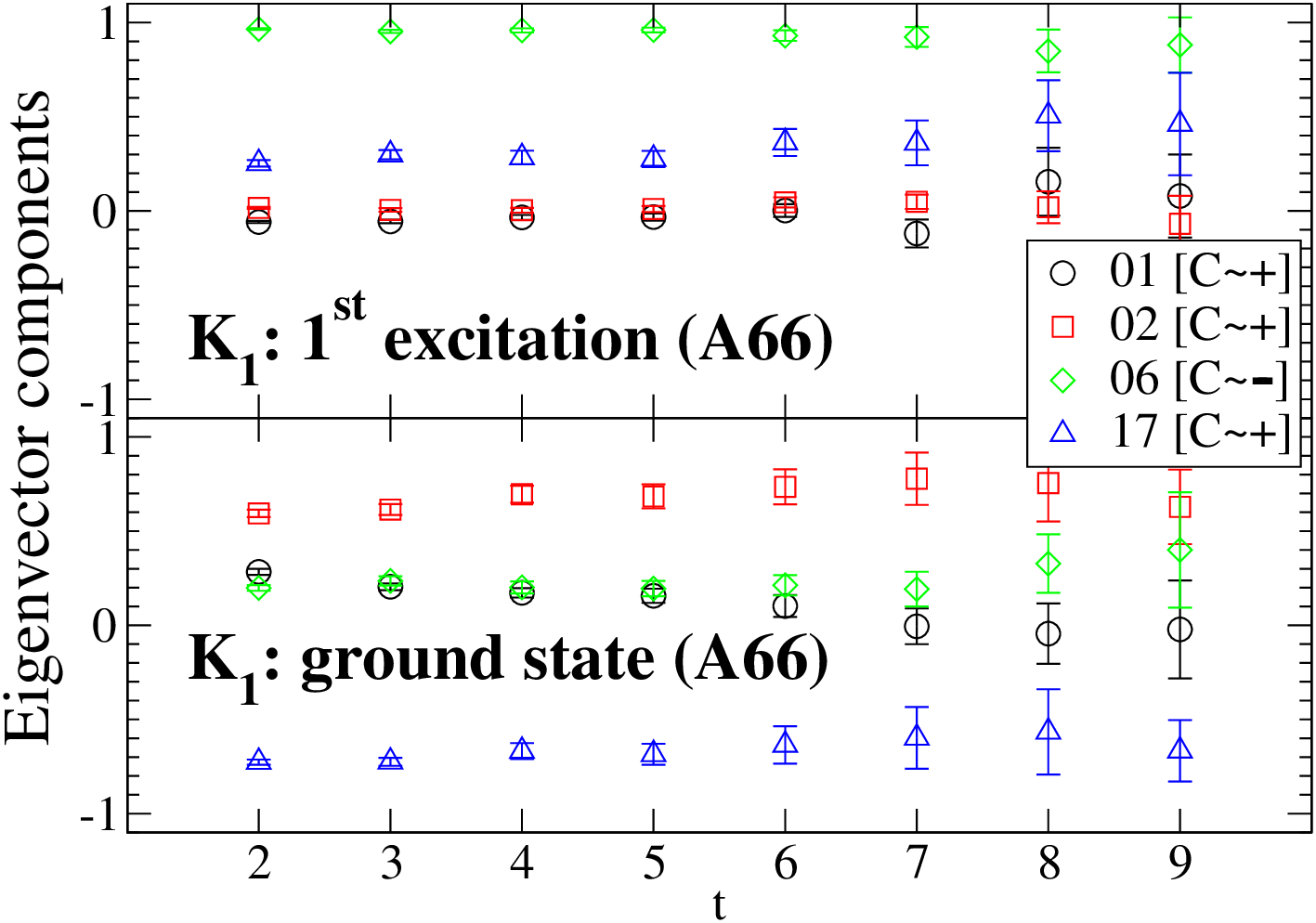}\\
\noindent\includegraphics[width=\columnwidth,clip]{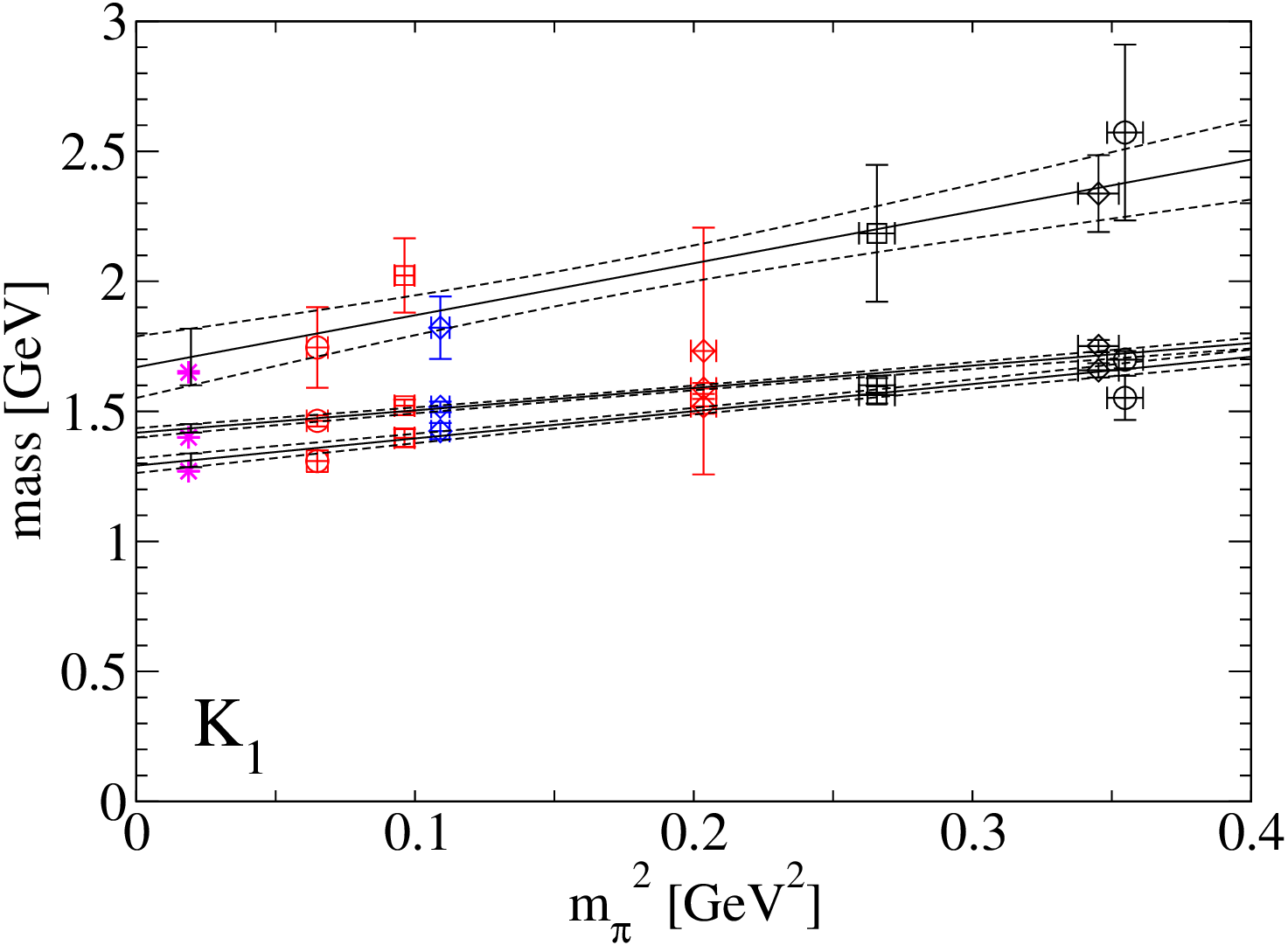}\
\caption{$1^{+}$ ($K_1$). Results for the energy levels are shown at the bottom.
The corresponding eigenvectors for the ground state and the first excitation for the lightest pion mass (A66) are shown on top.
Interpolators (1,2,17) have $[C\approx+]$, (6) has $[C\approx-]$.
Note the dominance of positive (negative) $C$-parity in the ground state (first excitation).
Note furthermore that there is some mixing in both states, which is allowed by the breaking of $C$-parity towards light pion masses.
At our largest pion masses, this mixing is suppressed strongly.}
\label{strange-mesons:1b}
\end{figure}

\begin{figure}
\noindent\includegraphics[width=\columnwidth,clip]{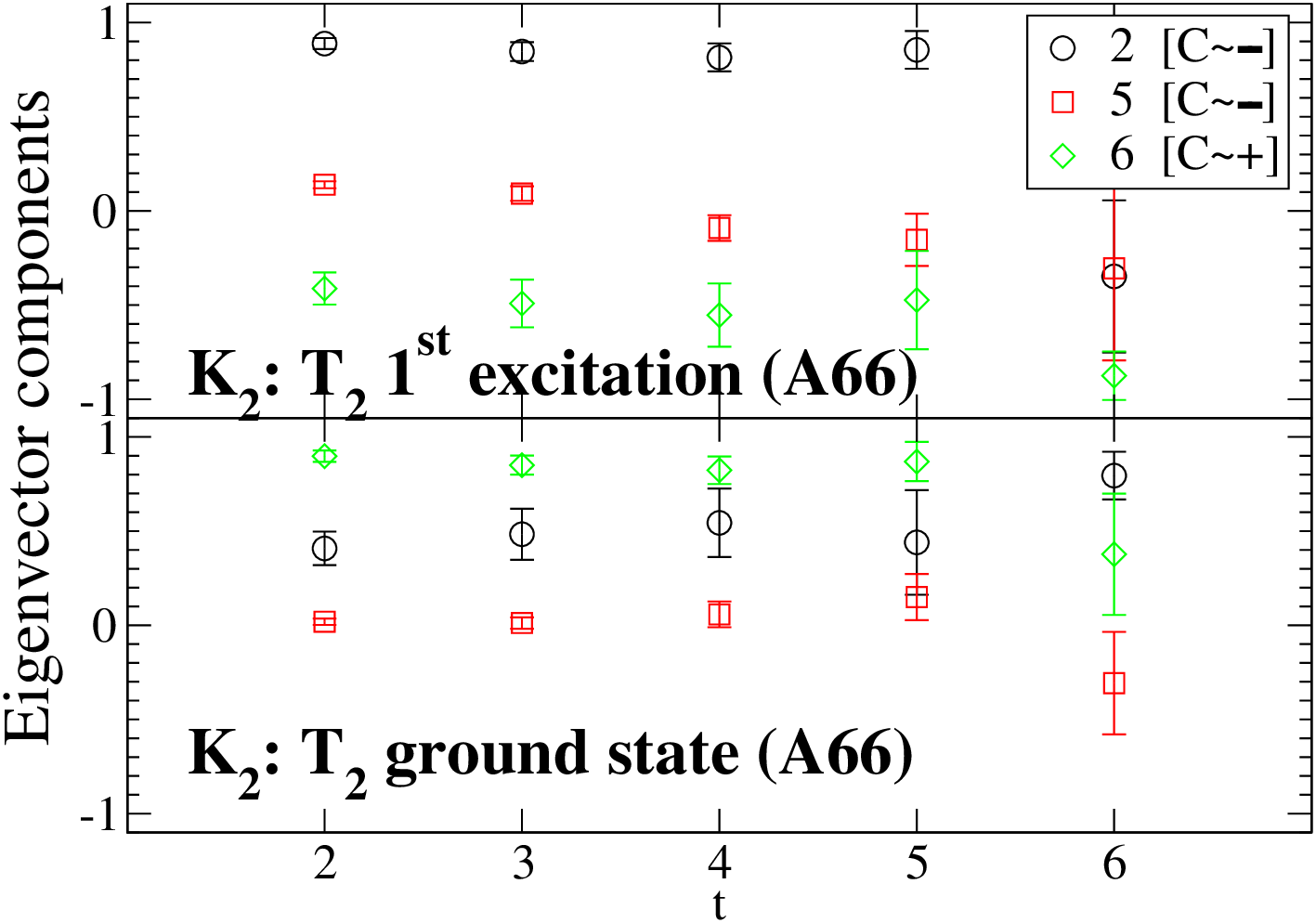}
\caption{
$(I=\frac{1}{2})\;\;2^{-}$ ($K_2$), representation $T_2$: 
The eigenvectors for the ground state and the first excitation for the lightest pion mass (A66) are shown.
Interpolators (2,5) have $[C\approx-]$, (6) has $[C\approx+]$.
Note the dominance of positive (negative) $C$-parity in the ground state (first excitation).
Note furthermore that there is significant mixing in both states, which is allowed by the breaking of $C$-parity towards light pion masses.
At our largest pion masses, this mixing is suppressed.
The mixing pattern is similar in representation $E$ (not shown).}
\label{strange-mesons:2a_vec}
\end{figure}

\begin{figure*}
\noindent\includegraphics[width=\columnwidth,clip]{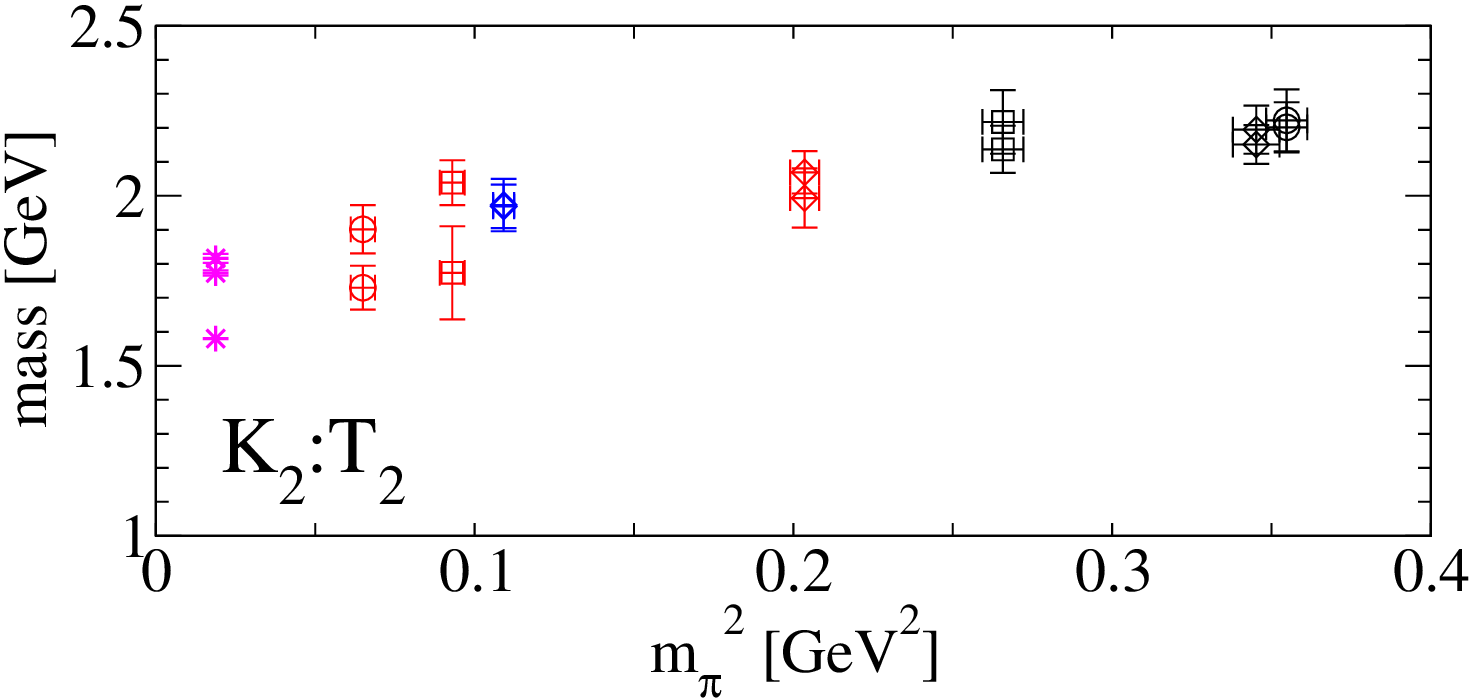}
\noindent\includegraphics[width=\columnwidth,clip]{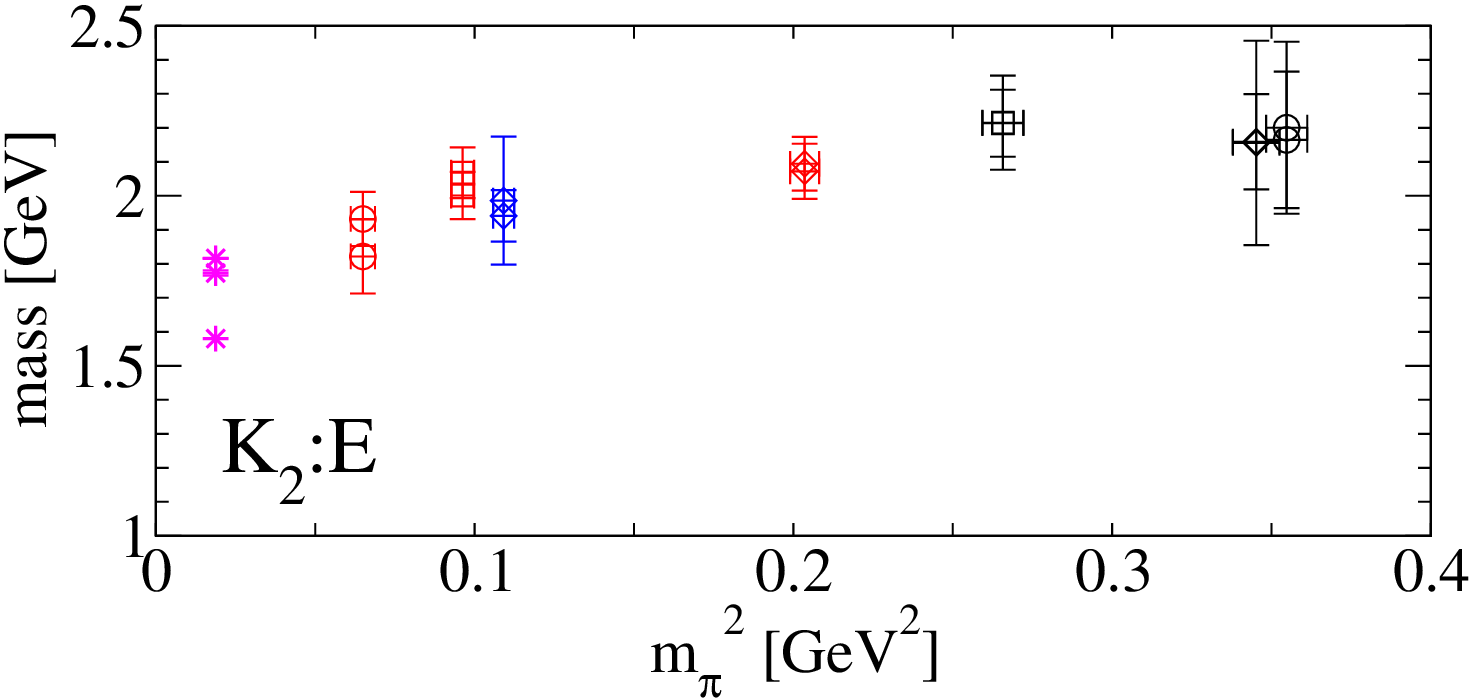}\\
\noindent\noindent\includegraphics[width=\columnwidth,clip]{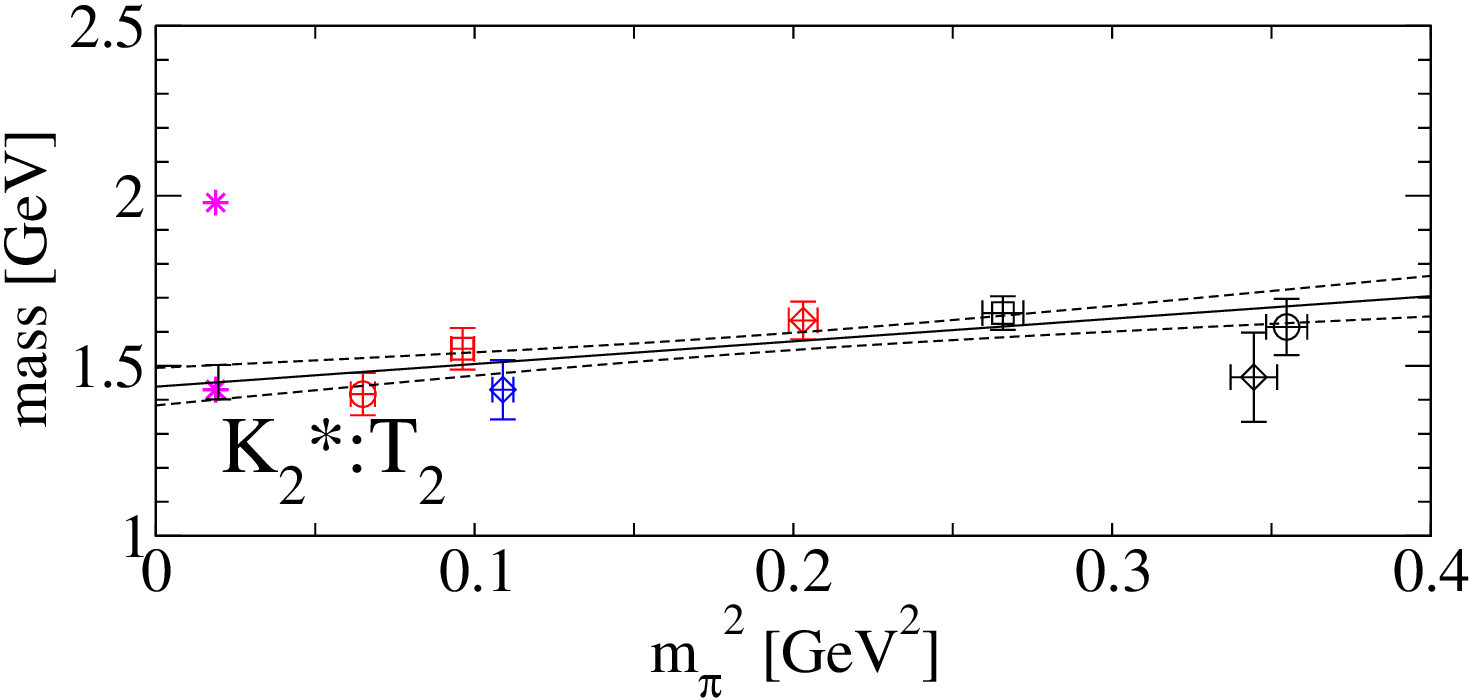}
\noindent\noindent\includegraphics[width=\columnwidth,clip]{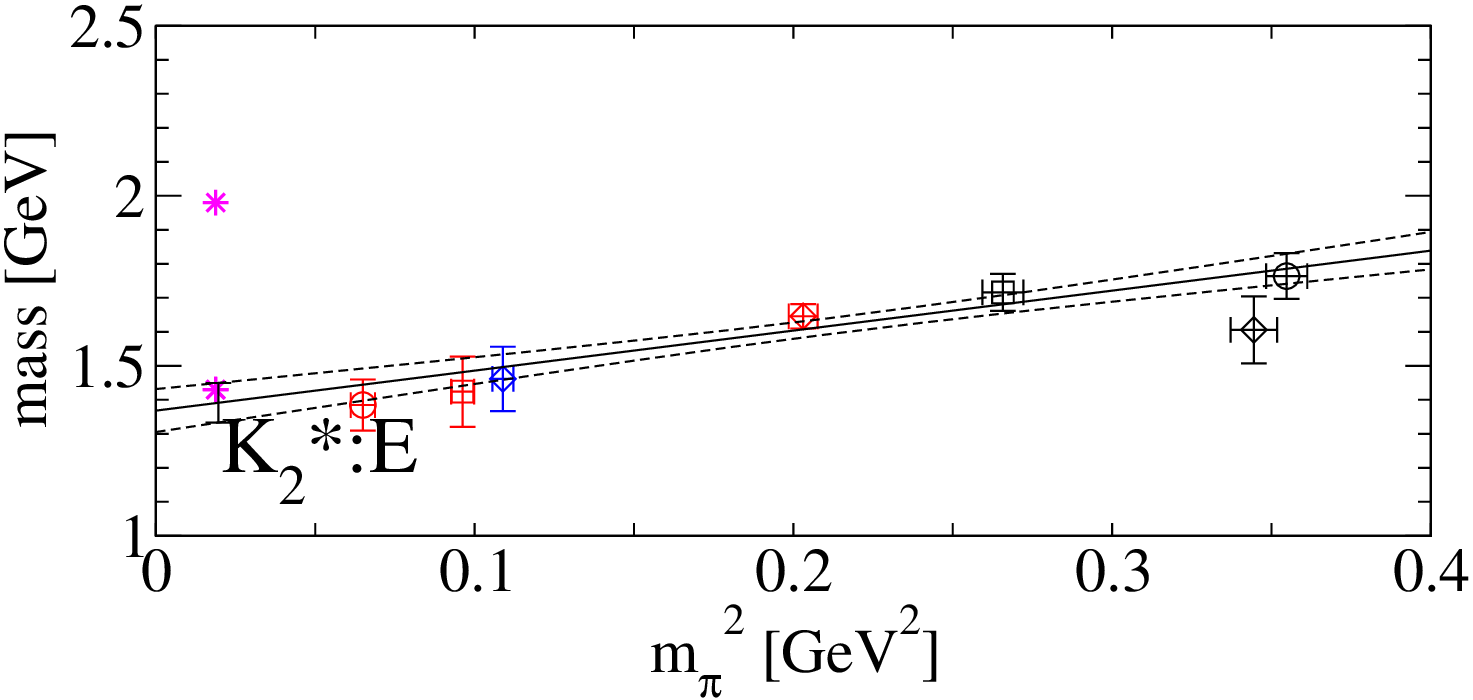}
\caption{Upper panels: $(I=\frac{1}{2})\;\;2^{-}$ ($K_2$) in both representations $T_2$ and $E$. 
Chiral fits are suppressed for clarity.\\
Lower panels: $(I=\frac{1}{2})\;\;2^{+}$ in both representations $T_2$ and $E$ (reliable signal only in $[C\approx +]$).
}
\label{strange-mesons:2a}
\end{figure*}

Finally, taking the set (1,8,9,12,16,20,21), both types of $C$-parities are included
in the variational method.  In this analysis, the three lowest states are dominated
by $[C\approx-]$ interpolators,  where even for the excitations the mixing is
compatible with zero.  A slight mixing is observed in ensemble A66, however, the
signal is very weak, and  the corresponding energy levels cannot be extracted
reliably.  One might wonder why we do not see a significant contribution of
$[C\approx+]$ interpolators to at least one of the excitations. A possible
interpretation is that the mixing is indeed weak in this channel at all simulated
pion masses  and that there is a further state, dominated by $[C\approx+]$, which is
not clearly identified in the full analysis.  The chiral extrapolations of the
excitations come out a bit high compared to the experimental $K^*(1410)$ and
$K^*(1680)$,  suggesting that simulations at smaller pion masses and with higher
statistics are necessary in order to reliably describe the mixing of different
$C$-parities and to be able to obtain the $K^*(1410)$.

\myparagraph{(I=\frac{1}{2})\;\;1^{+} (K_1)}
Looking at the experimental states in the corresponding light meson channels
$a_1(1260)$, $b_1(1235)$ and $a_1(1640)$, mixing is expected already for the
lowest states $K_1(1270)$, $K_1(1400)$ and $K_1(1650)$.

Employing pure $[C\approx+]$ sets of interpolators, 
the chiral extrapolation of the ground
state ends up between the $K_1(1270)$ and the $K_1(1400)$. The first
excitation hits the $K_1(1650)$ within error bars.
From pure $[C\approx-]$ interpolators only a ground state can be extracted,
the chiral extrapolation of which agrees with the $K_1(1400)$.

Allowing for both types of $C$-parity, three states can be extracted when
the set of interpolators is optimized in each ensemble. 
The chiral extrapolations are
compatible with $K_1(1270)$, $K_1(1400)$ and $K_1(1650)$ 
(see Fig.~\ref{strange-mesons:1b}). 
Since the splitting of
$K_1(1270)$ and $K_1(1400)$ is rather small, it is hard to make a statement
about its increase towards smaller pion masses. 
(Notice that an increased splitting is observed when mixing both charged conjugations for the analogous mesons in the charmed meson sector \cite{Mohler:2011ke}.) This is worsened by the
fluctuation of the plateau points. However, the eigenvectors indeed show stronger
mixing approaching the physical point (see Fig.~\ref{strange-mesons:1b}), 
which is usually accompanied by a more
pronounced splitting. At simulated pion masses, $K_1(1270)$ and $K_1(1650)$
are dominated by $[C\approx+]$, $K_1(1400)$ by $[C\approx-]$ interpolators.
Our results confirm the existence of $K_1(1650)$ (omitted from the summary
table of \cite{Nakamura:2010zzi}), which is dominated by positive $C$-parity 
in our analysis.

\begin{figure}
\noindent\includegraphics[width=\columnwidth,clip]{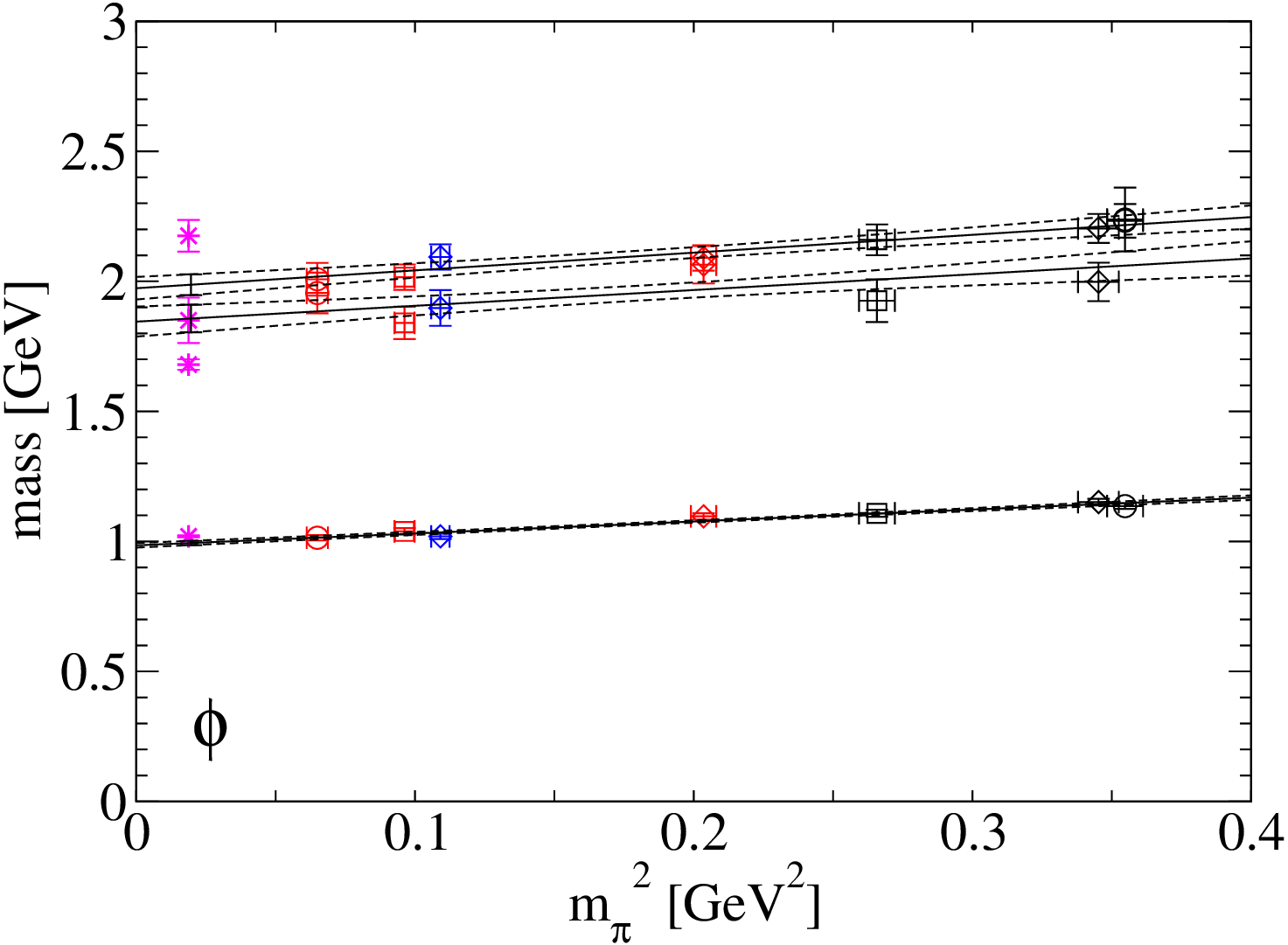}\\
\noindent\includegraphics[width=\columnwidth,clip]{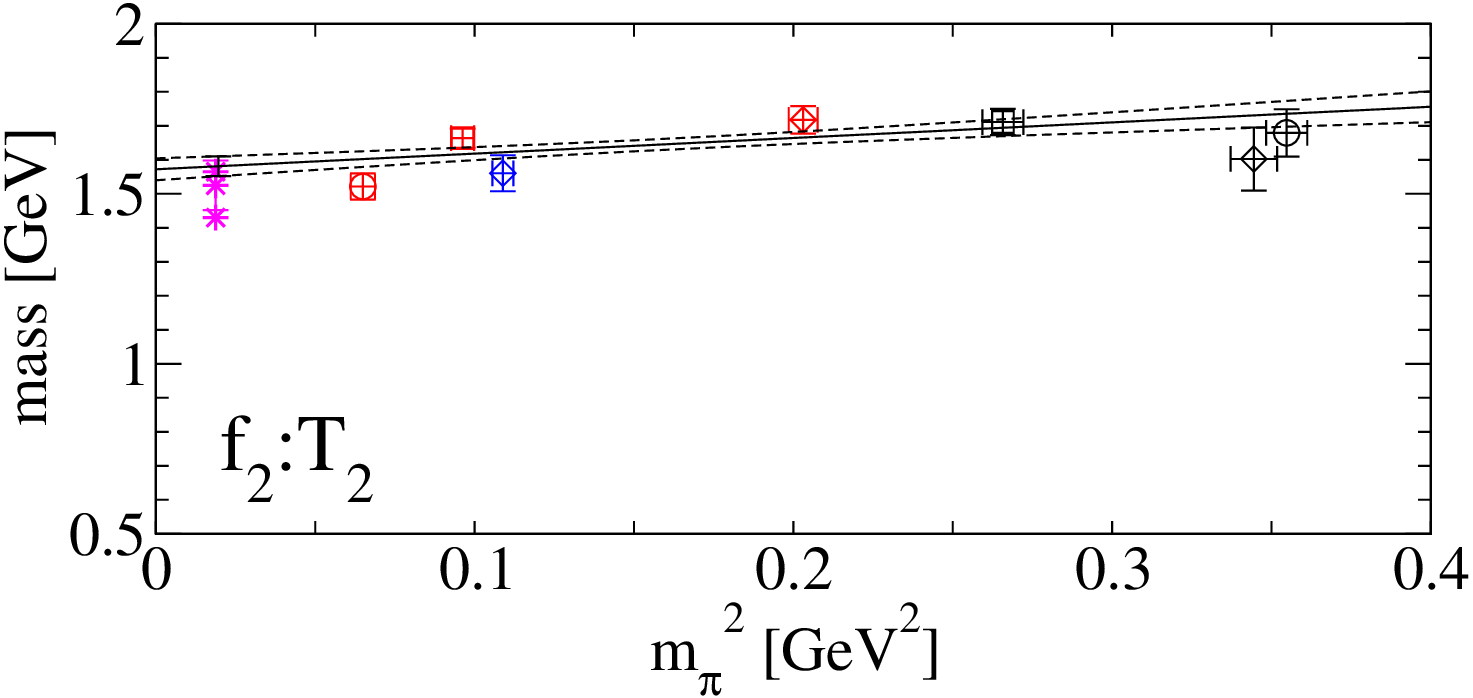}\\
\noindent\includegraphics[width=\columnwidth,clip]{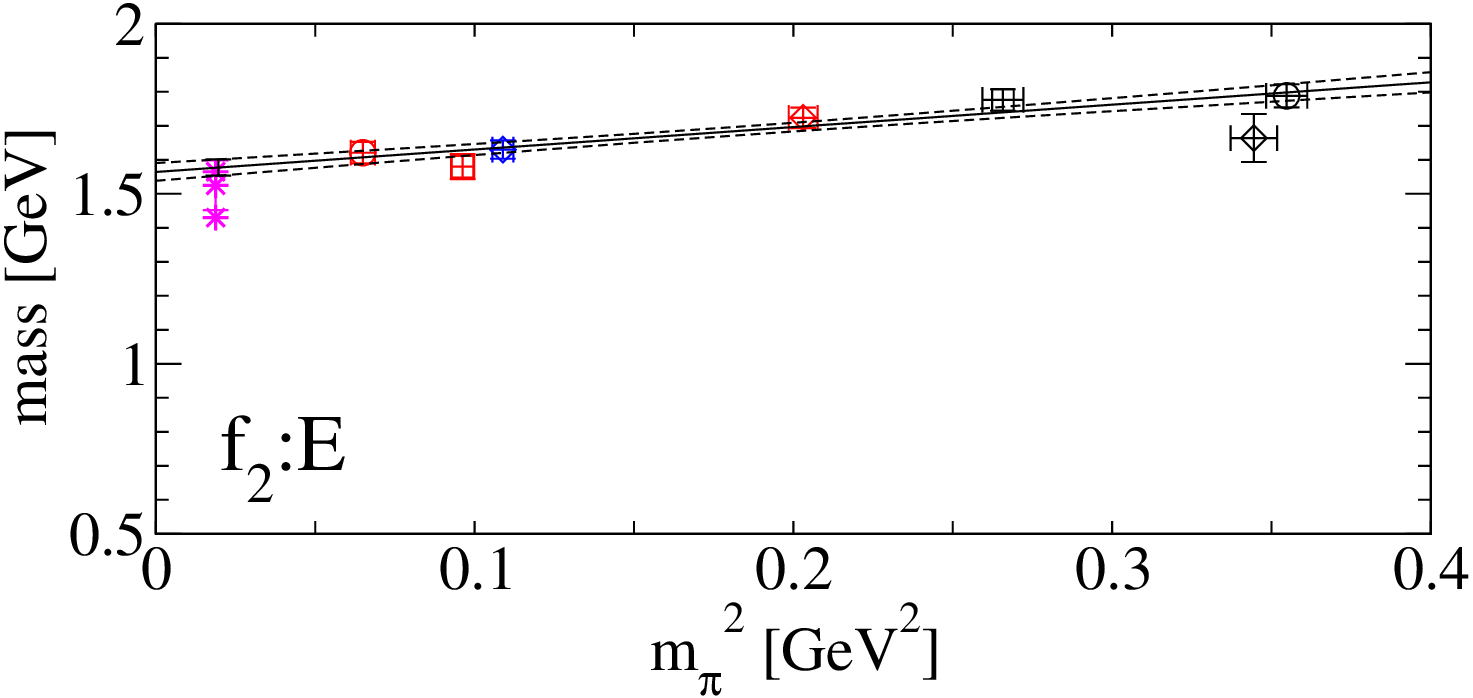}\\
\caption{(top) $(s \overline s) 1^{--}$; (middle and bottom)   $(s \overline
s) 2^{++}$.
}
\label{isoscalar-mesons}
\end{figure}

\begin{figure}
\noindent\includegraphics[width=\columnwidth,clip]{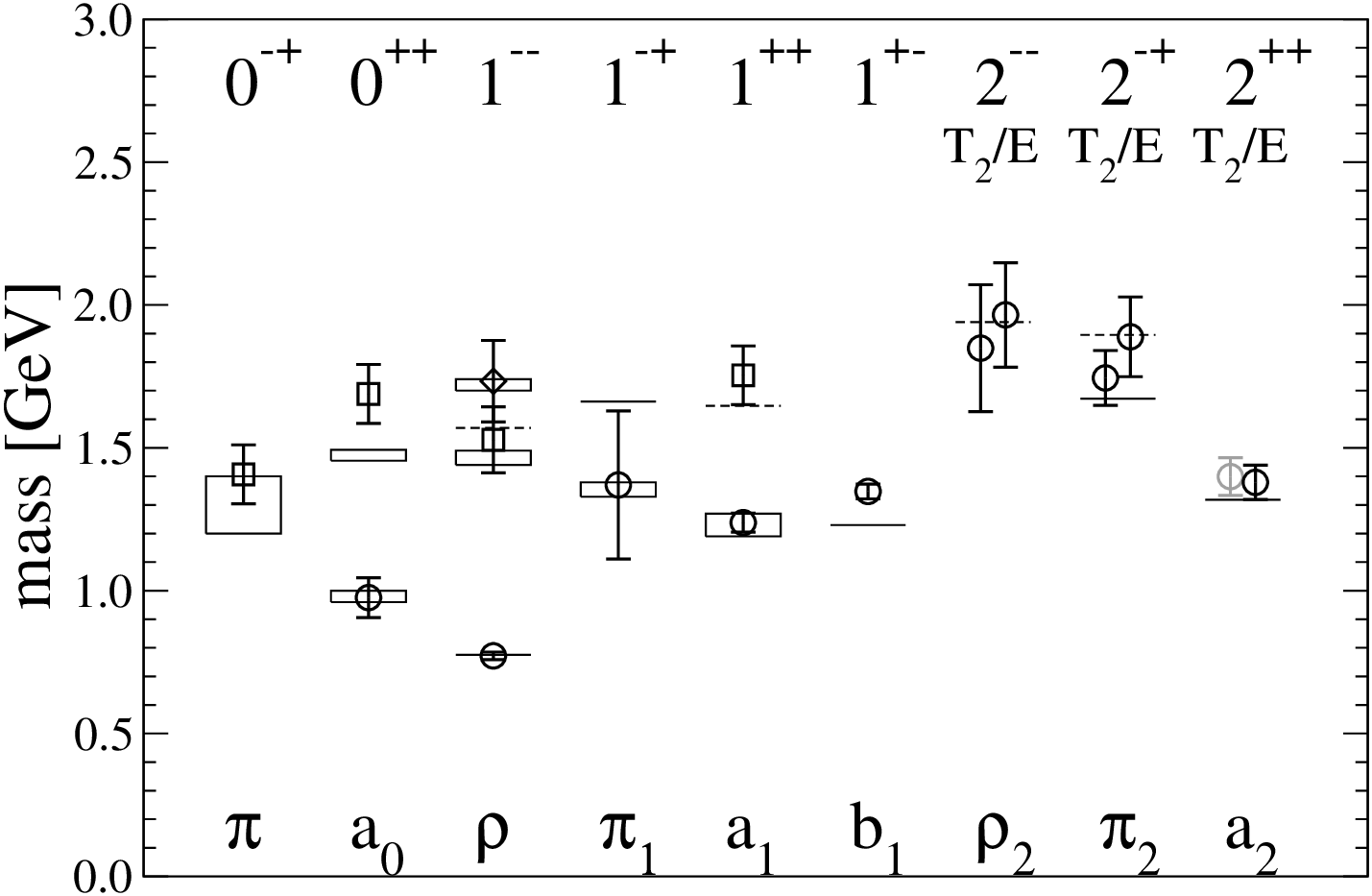}\\
\caption{Results for the isovector light meson masses.
All values are obtained by chiral extrapolation linear in the pion mass squared.
Horizontal lines or boxes represent experimentally known states, 
dashed lines indicate poor evidence, according to \cite{Nakamura:2010zzi}.
The statistical uncertainty of our results is indicated by bands of 1 $\sigma$, 
that of the experimental values by boxes of 1 $\sigma$.
In case of spin 2 mesons, results for $T_2$ and $E$ are shown side by side.
Grey symbols denote a poor $\chi^2$/d.o.f.~of the chiral fits (see Tab.~\ref{tab:chi2lightmesons}).
}
\label{fig:mesons_summary}
\end{figure}

\begin{figure}
\noindent\includegraphics[width=\columnwidth,clip]{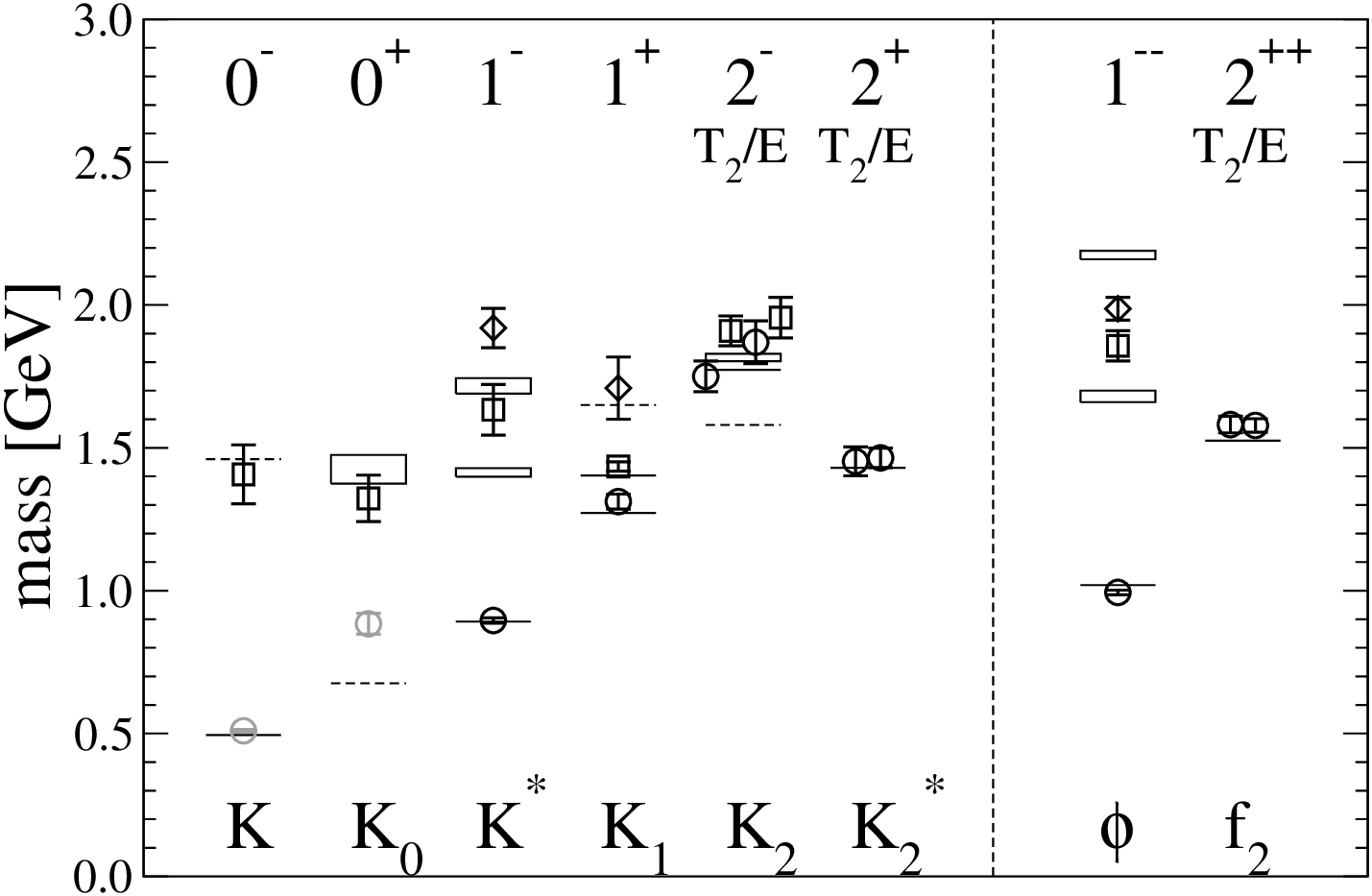}
\caption{Same as Fig.\ref{fig:mesons_summary}, but for strange mesons (left) and isoscalars (right).
The strange quarks are implemented in the partial quenching approximation.
The isoscalars additionally suffer from neglected disconnected diagrams.
In the isoscalar sector we show experimental results only where they
have a dominant $s\overline{s}$ content \cite{Nakamura:2010zzi}.
Grey symbols denote a poor $\chi^2$/d.o.f.~of the chiral fits (see Tabs.\ref{tab:chi2strangemesons} and \ref{tab:chi2isoscalarmesons}).
}
\label{fig:strangemesons_summary}
\end{figure}

\subsection{Tensors}

\myparagraph{(I=\frac{1}{2})\;\;2^{-} (K_2)}
In the spin 2 channels, investigation of the mixing becomes more complicated,
since the signal is often weak already for the ground state. From the light
meson states $\pi_2(1670)$, $\pi_2(1880)$ and the (not established)
$\rho_2(1940)$,  one could expect a dominance of $[C\approx+]$ interpolators
in the ground state. So far, $K_2(1580)$ is omitted from the summary table of
\cite{Nakamura:2010zzi}, the lowest established states in this channel are
$K_2(1770)$ and $K_2(1820)$.

Restricting the basis to negative $C$-parity, we use interpolator (2) as in the corresponding light channel. 
In both $T_2$ and $E$, the
chiral extrapolation is compatible with $K_2(1770)$ and $K_2(1820)$. 
For positive $C$-parity, using interpolator (6) in $T_2$ and (8) in $E$, the chiral extrapolations are
again compatible with $K_2(1770)$ and $K_2(1820)$.

To take into account both $C$-parities, the set (2,5,6) (resp.~(3,4,5,6) in C72) is
chosen in $T_2$ and (2,5,8) in $E$.  The two lowest eigenvalues are very close and
have to be sorted according to the eigenvectors.  The eigenvectors of $T_2$ are
shown in Fig.~\ref{strange-mesons:2a_vec}. We observe that the ground (excited)
state is dominated by positive (negative) $C$-parity. However, there is significant
mixing in both states, which appears to be the strongest mixing  of all channels
considered. Strong mixing is also observed in representation $E$.
The chiral extrapolations are compatible with the experimentally established
$K_2(1770)$ and $K_2(1820)$ (see upper panels of Fig.~\ref{strange-mesons:2a}) and do not confirm
the $K_2(1580)$, which is omitted from the summary table of
\cite{Nakamura:2010zzi}.  However, increasing mixing towards lighter pion masses
could still change the slope of the chiral extrapolation.

\myparagraph{(I=\frac{1}{2})\;\;2^{+} (K_2^*)}
No experimental state is known in the light-quark $2^{+{-}}$ channel. In the light
$2^{+{+}}$ channel, the lowest states are $a_2(1320)$, $a_2(1700)$ and
$a_2(1950)$, of which the latter two are not established. In the strange 
$2^{+{-}}$ channel the lowest experimental states are $K_2^*(1430)$ and the
(not established) $K_2^*(1980)$. 

The signal of negative $C$-parity interpolators is weak here, 
thus we restrict our analysis to positive $C$-parity interpolators. 
Interpolator (2) 
(Table \ref{tab:interpols_2+T_2}) is used in $T_2$ 
and interpolator (2) (Table \ref{tab:interpols_2+E}) in $E$ 
to extract a ground state mass. In
both lattice channels, the chiral extrapolation hits the experimental
$K_2^*(1430)$ nicely (see lower panels of Fig.~\ref{strange-mesons:2a}).

\subsection{Isoscalar light mesons}

\myparagraph{1^{--} (\phi)}
In principle, correlation functions of isoscalar mesons include connected and
disconnected diagrams. The low lying isoscalar $\phi$ mesons decay mainly into
kaons, thus one expects that these states are dominated by strange quarks 
(Zweig rule).  Since disconnected diagrams are dominated by loops
of light sea quarks, it is reasonable to assume that the $\phi$ mesons are
dominated by connected (strange) diagrams.  We extract $\phi$ meson masses
evaluating only these connected
diagrams, albeit with the systematic error of
neglecting the disconnected diagrams. We use the same set of operators as in
the light isovector $1^{-{-}}$ ($\rho$) channel (Sec. \ref{light-vectors}) 
to extract three energy
levels. 

The ground state mass extrapolates to a value very close to the experimental $\phi(1020)$ mass 
(see Fig.~\ref{isoscalar-mesons}),
which confirms our choice of the strange quark mass parameters. The
extrapolation of the excited states ends up significantly higher than the
experimental $\phi(1680)$. Since the first
excitation $\rho(1450)$ in the light isovector channel is reproduced nicely,
one may conclude that the neglected disconnected diagrams play a more
important role for the $\phi(1680)$ compared to the $\phi(1020)$. The lattice
irreducible representation $T_1$ couples to continuum spins 1 and 3 (among
others). This is why we also indicate the possible spin 3 state $\phi_3(1850)$
in the figure. This state is hit by the extrapolations of the first and second
excitation. However, all our interpolators in this channel have a na\"ive
continuum limit of spin 1. One of these two levels may bend down if
disconnected diagrams are included.

\myparagraph{2^{++} (f_2)}
As in the $\phi$ meson channel, the experimental decay channels of the
isoscalar light meson $f_2$ suggest dominance of connected diagrams. We use
the same interpolators as in the isovector $2^{+{+}}$ ($a_2$) channel. The
results of $T_2$ and $E$ agree (see Fig.~\ref{isoscalar-mesons}), but their
chiral extrapolations are in better
agreement with the $f_2'(1525)$ than with the $f_2(1430)$. The latter needs
confirmation and is not listed in the summary table of
\cite{Nakamura:2010zzi}. It is unclear if inclusion of the
neglected disconnected diagrams would yield the $f_2(1430)$ or if the ground
state of the theory is the established $f_2'(1525)$.

\section{Conclusions}

We presented results for the light and strange meson spectrum from two dynamical
Chirally Improved quarks. Seven ensembles with pion masses between 250 and 600
MeV were analyzed with the  variational method in order to extract energy levels
for ground and excited states. In addition to dynamical light quarks we also
included strange quarks within the partially quenched approximation, 
fitting the strange quark mass by requiring
the correct $\Omega(1672)$ mass.

Figure \ref{fig:mesons_summary} shows our chirally extrapolated results for the
spectrum of light mesons compared to experimental values from
\cite{Nakamura:2010zzi}. Figure \ref{fig:strangemesons_summary} contains a
similar plot for strange mesons (left panel) and isoscalars (right panel). The
results  are in general in good agreement with experiment.  For the strange
mesons the good agreement for the ground states in the kaon, the $K^\star$ and
$\phi$ meson channels suggest that these observables are well-reproduced in the
partially quenched approximation and confirm our choice of strange quark mass
parameter. As discussed in more detail in Sections \ref{lightmesons} and
\ref{strangemesons}, we do not see any clear indications of scattering states,
which probably show only little overlap with the one-particle interpolators used
in this work. Exceptions are the strange $0^+$ channel and the light isovector $0^+$
channel at small quark masses, where our signal is also consistent with
a two-particle scattering state.

The strange meson channels $1^-$, $1^+$ and $2^-$ have been investigated with
respect to their approximate $C$-parity. In the $1^-$ channel, the three lowest states
seem to be dominated by negative $C$-parity, while positive $C$-parity was shown
to contribute to a state in the vicinity of the second excitation. The $1^+$
channel shows some mixing of different $C$-parity towards light pion masses, and
the low lying spectrum seems to contain states with alternating $C$-parity
dominance. The $2^-$ channel shows strong mixing towards light pion masses and
the ground state (first excitation) is dominated by positive (negative)
$C$-parity.

For our lightest three pion masses finite size effects may play a non-negligible
role and their influence on our results deserves further attention. A study on
larger volumes is in progress and we will investigate this source of possible
systematic errors in the near future. The larger volume will also be used for
the spectroscopy of low-lying baryon states, where finite volume effects are
expected to be more pronounced.

\acknowledgments
We would like to thank Christof Gattringer and Leonid Y.~Glozman for valuable
discussions. The calculations have been performed on the SGI Altix 4700 of the
Leibniz-Rechenzentrum Munich and on local clusters at ZID at the University of
Graz. We thank these institutions for providing	 support. M.L.~has been
supported by Austrian Science Fund (FWF: DK W1203-N16) and by EU FP7 project HadronPhysics2. 
D.M.~acknowledges support by Natural Sciences
and Engineering Research Council of Canada (NSERC) and G.P.E., M.L.~and
A.S.~acknowledge support by the DFG project SFB/TR-55.

\begin{appendix}
\section{Tables of interpolators}
\label{interpolators}

In the tables for meson interpolators (Table \ref{tab:interpols_0-} to
\ref{tab:interpols_2+T_2}),  the two quark fields are labeled by $a$ and $b$.
These are placeholders for light ($u,d$) or strange ($s$) quarks.  The indices
$n$, $w$ and $\partial_i$ correspond to the smearings narrow, wide and
derivative, respectively.  $\gamma_i$ are the spatial Dirac matrices,
$\gamma_t$ is the Dirac matrix in time direction. $\epsilon_{ijk}$ is the
Levi-Civita symbol,
$Q_{ijk}$ are Clebsch-Gordon coefficients, where all elements are zero except
$Q_{111}=\frac{1}{\sqrt{2}}$, 
$Q_{122}=-\frac{1}{\sqrt{2}}$, 
$Q_{211}=-\frac{1}{\sqrt{6}}$, 
$Q_{222}=-\frac{1}{\sqrt{6}}$ and  
$Q_{233}=\frac{2}{\sqrt{6}}$.

\begin{table}[!]
\begin{ruledtabular}
\begin{tabular}{lcc}
\#$_{0^-}$ & Interpolator & $C$\\
\hline
 1  &  $\ov{a}_n \gamma_5 b_n$ & $+$ \\
 2  &  $\ov{a}_n \gamma_5 b_w + \ov{a}_w \gamma_5 b_n$ & $+$ \\
 3  &  $\ov{a}_n \gamma_5 b_w - \ov{a}_w \gamma_5 b_n$ & $-$ \\
 4  &  $\ov{a}_w \gamma_5 b_w$ & $+$ \\[0.3cm]
 5  &  $\ov{a}_n \gamma_t\gamma_5 b_n$ & $+$ \\
 6  &  $\ov{a}_n \gamma_t\gamma_5 b_w + \ov{a}_w \gamma_t\gamma_5 b_n$ & $+$ \\
 7  &  $\ov{a}_n \gamma_t\gamma_5 b_w - \ov{a}_w \gamma_t\gamma_5 b_n$ & $-$ \\
 8  &  $\ov{a}_w \gamma_t\gamma_5 b_w$ & $+$ \\[0.3cm]
 9  &  $\ov{a}_{\partial_i} \gamma_i\gamma_5 b_n + \ov{a}_n \gamma_i\gamma_5 b_{\partial_i}$ & $+$ \\
10  &  $\ov{a}_{\partial_i} \gamma_i\gamma_5 b_n - \ov{a}_n \gamma_i\gamma_5 b_{\partial_i}$ & $-$ \\
11  &  $\ov{a}_{\partial_i} \gamma_i\gamma_5 b_w + \ov{a}_w \gamma_i\gamma_5 b_{\partial_i}$ & $+$ \\
12  &  $\ov{a}_{\partial_i} \gamma_i\gamma_5 b_w - \ov{a}_w \gamma_i\gamma_5 b_{\partial_i}$ & $-$ \\[0.3cm]
13  &  $\ov{a}_{\partial_i} \gamma_i\gamma_t\gamma_5 b_n + \ov{a}_n \gamma_i\gamma_t\gamma_5 b_{\partial_i}$ & $-$ \\
14  &  $\ov{a}_{\partial_i} \gamma_i\gamma_t\gamma_5 b_n - \ov{a}_n \gamma_i\gamma_t\gamma_5 b_{\partial_i}$ & $+$ \\
15  &  $\ov{a}_{\partial_i} \gamma_i\gamma_t\gamma_5 b_w + \ov{a}_w \gamma_i\gamma_t\gamma_5 b_{\partial_i}$ & $-$ \\
16  &  $\ov{a}_{\partial_i} \gamma_i\gamma_t\gamma_5 b_w - \ov{a}_w \gamma_i\gamma_t\gamma_5 b_{\partial_i}$ & $+$ \\[0.3cm]
17  &  $\ov a_{\partial_i} \gamma_5 b_{\partial_i}$ & $+$ \\[0.3cm]
18  &  $\ov a_{\partial_i} \gamma_t\gamma_5 b_{\partial_i}$ & $+$ \\[0.3cm]
\end{tabular}
\end{ruledtabular}
\caption{Meson interpolators for $J^P=0^-$. 
The first row shows the number, the second shows the explicit form of the
interpolator. In the
last column the $C$ parity is given, which is only an approximate quantum
number in the case of differing quark masses.  Interpolators with different quark field smearings and similar Dirac structure are grouped and these groups separated by
white space.
}
\label{tab:interpols_0-}
\end{table}


\begin{table}[!]
\begin{ruledtabular}
\begin{tabular}{lcc}
\#$_{0^+}$ & interpolator(s) & $C$ parity\\
\hline
 1    &  $\ov{a}_n b_n$                                      &$+$\\
 2    &  $\ov{a}_n b_w + \ov{a}_w b_n$                              &$+$\\
 3    &  $\ov{a}_n b_w - \ov{a}_w b_n$                              &$-$\\
 4    &  $\ov{a}_w b_w$                                      &$+$\\[0.3cm]
 5    &  $\ov{a}_{\partial_i} \gamma_i b_n + \ov{a}_n \gamma_i b_{\partial_i}$          &$-$\\
 6    &  $\ov{a}_{\partial_i} \gamma_i b_n - \ov{a}_n \gamma_i b_{\partial_i}$          &$+$\\
 7    &  $\ov{a}_{\partial_i} \gamma_i b_w + \ov{a}_w \gamma_i b_{\partial_i}$          &$-$\\
 8    &  $\ov{a}_{\partial_i} \gamma_i b_w - \ov{a}_w \gamma_i b_{\partial_i}$          &$+$\\[0.3cm]
 9    &  $\ov{a}_{\partial_i} \gamma_i\gamma_t b_n + \ov{a}_n \gamma_i\gamma_t b_{\partial_i}$  &$-$\\
10    &  $\ov{a}_{\partial_i} \gamma_i\gamma_t b_n - \ov{a}_n \gamma_i\gamma_t b_{\partial_i}$  &$+$\\
11    &  $\ov{a}_{\partial_i} \gamma_i\gamma_t b_w + \ov{a}_w \gamma_i\gamma_t b_{\partial_i}$  &$-$\\
12    &  $\ov{a}_{\partial_i} \gamma_i\gamma_t b_w - \ov{a}_w \gamma_i\gamma_t b_{\partial_i}$  &$+$\\[0.3cm]
13    &  $\ov{a}_{\partial_i} b_{\partial_i}$                           &$+$\\
\end{tabular}
\end{ruledtabular}
\caption{Same as Tab.\ \ref{tab:interpols_0-}, now for $J^P=0^+$.}
\end{table}


\begin{table}[!]
\begin{ruledtabular}
\begin{tabular}{lcc}
\#$_{1^-}$ & interpolator(s) & $C$\\
\hline
 1    &  $\ov{a}_n \gamma_k b_n$                                                        &$-$\\
 2    &  $\ov{a}_n \gamma_k b_w + \ov{a}_w \gamma_k b_n$                                            &$-$\\
 3    &  $\ov{a}_n \gamma_k b_w - \ov{a}_w \gamma_k b_n$                                            &$+$\\
 4    &  $\ov{a}_w \gamma_k b_w$                                                        &$-$\\[0.3cm]
 5    &  $\ov{a}_n \gamma_k\gamma_t b_n$                                                    &$-$\\
 6    &  $\ov{a}_n \gamma_k\gamma_t b_w + \ov{a}_w \gamma_k\gamma_t b_n$                                    &$-$\\
 7    &  $\ov{a}_n \gamma_k\gamma_t b_w - \ov{a}_w \gamma_k\gamma_t b_n$                                    &$+$\\
 8    &  $\ov{a}_w \gamma_k\gamma_t b_w$                                                    &$-$\\[0.3cm]
 9    &  $\ov{a}_{\partial_k} b_n + \ov{a}_n b_{\partial_k}$                                          &$+$\\
10    &  $\ov{a}_{\partial_k} b_n - \ov{a}_n b_{\partial_k}$                                          &$-$\\
11    &  $\ov{a}_{\partial_k} b_w + \ov{a}_w b_{\partial_k}$                                          &$+$\\
12    &  $\ov{a}_{\partial_k} b_w - \ov{a}_w b_{\partial_k}$                                          &$-$\\[0.3cm]
13    &  $\ov{a}_{\partial_k} \gamma_t b_n + \ov{a}_n \gamma_t b_{\partial_k}$                                 &$-$\\
14    &  $\ov{a}_{\partial_k} \gamma_t b_n - \ov{a}_n \gamma_t b_{\partial_k}$                                 &$+$\\
15    &  $\ov{a}_{\partial_k} \gamma_t b_w + \ov{a}_w \gamma_t b_{\partial_k}$                                 &$-$\\
16    &  $\ov{a}_{\partial_k} \gamma_t b_w - \ov{a}_w \gamma_t b_{\partial_k}$                                 &$+$\\[0.3cm]
17    &  $\ov{a}_{\partial_i} \gamma_k b_{\partial_i}$                                             &$-$\\[0.3cm]
18    &  $\ov{a}_{\partial_i} \gamma_k\gamma_t b_{\partial_i}$                                         &$-$\\[0.3cm]
19    &  $\ov{a}_{\partial_k} \epsilon_{ijk} \gamma_j\gamma_5 b_n + \ov{a}_n \epsilon_{ijk} \gamma_j\gamma_5 b_{\partial_k}$          &$+$\\
20    &  $\ov{a}_{\partial_k} \epsilon_{ijk} \gamma_j\gamma_5 b_n - \ov{a}_n \epsilon_{ijk} \gamma_j\gamma_5 b_{\partial_k}$          &$-$\\
21    &  $\ov{a}_{\partial_k} \epsilon_{ijk} \gamma_j\gamma_5 b_w + \ov{a}_w \epsilon_{ijk} \gamma_j\gamma_5 b_{\partial_k}$          &$+$\\
22    &  $\ov{a}_{\partial_k} \epsilon_{ijk} \gamma_j\gamma_5 b_w - \ov{a}_w \epsilon_{ijk} \gamma_j\gamma_5 b_{\partial_k}$          &$-$\\[0.3cm]
23    &  $\ov{a}_{\partial_k} \epsilon_{ijk}\gamma_j\gamma_t\gamma_5 b_n  + \ov{a}_n \epsilon_{ijk}\gamma_j\gamma_t\gamma_5 b_{\partial_k}$ &$-$\\
24    &  $\ov{a}_{\partial_k} \epsilon_{ijk}\gamma_j\gamma_t\gamma_5 b_n  - \ov{a}_n \epsilon_{ijk}\gamma_j\gamma_t\gamma_5 b_{\partial_k}$ &$+$\\
25    &  $\ov{a}_{\partial_k} \epsilon_{ijk}\gamma_j\gamma_t\gamma_5 b_w  + \ov{a}_w \epsilon_{ijk}\gamma_j\gamma_t\gamma_5 b_{\partial_k}$ &$-$\\
26    &  $\ov{a}_{\partial_k} \epsilon_{ijk}\gamma_j\gamma_t\gamma_5 b_w  - \ov{a}_w \epsilon_{ijk}\gamma_j\gamma_t\gamma_5 b_{\partial_k}$ &$+$\\
\end{tabular}
\end{ruledtabular}
\caption{Same as Tab.\ \ref{tab:interpols_0-}, now for $J^P=1^-$.}
\label{tab:rho_ops}
\end{table}

\begin{table}[!]
\begin{ruledtabular}
\begin{tabular}{lcc}
\#$_{1^+}$ & interpolator(s) & $C$\\
\hline
 1    &  $\ov{a}_n \gamma_k \gamma_5 b_n$                                         &$+$\\
 2    &  $\ov{a}_n \gamma_k \gamma_5 b_w+\ov{a}_w \gamma_k \gamma_5 b_n$                         &$+$\\
 3    &  $\ov{a}_n \gamma_k \gamma_5 b_w-\ov{a}_w \gamma_k \gamma_5 b_n$                         &$-$\\
 4    &  $\ov{a}_w \gamma_k \gamma_5 b_w$                                         &$+$\\[0.3cm]
 5    &  $\ov{a}_{\partial_k} \gamma_5 b_n+\ov{a}_n \gamma_5 b_{\partial_k}$                       &$+$\\
 6    &  $\ov{a}_{\partial_k} \gamma_5 b_n-\ov{a}_n \gamma_5 b_{\partial_k}$                       &$-$\\
 7    &  $\ov{a}_{\partial_k} \gamma_5 b_w+\ov{a}_w \gamma_5 b_{\partial_k}$                       &$+$\\
 8    &  $\ov{a}_{\partial_k} \gamma_5 b_w-\ov{a}_w \gamma_5 b_{\partial_k}$                       &$-$\\[0.3cm]
 9    &  $\ov{a}_{\partial_k} \gamma_t \gamma_5 b_n+\ov{a}_n \gamma_t \gamma_5 b_{\partial_k}$              &$+$\\
10    &  $\ov{a}_{\partial_k} \gamma_t \gamma_5 b_n-\ov{a}_n \gamma_t \gamma_5 b_{\partial_k}$              &$-$\\
11    &  $\ov{a}_{\partial_k} \gamma_t \gamma_5 b_w+\ov{a}_w \gamma_t \gamma_5 b_{\partial_k}$              &$+$\\
12    &  $\ov{a}_{\partial_k} \gamma_t \gamma_5 b_w-\ov{a}_w \gamma_t \gamma_5 b_{\partial_k}$              &$-$\\[0.3cm]
13    &  $\ov{a}_{\partial_i} \gamma_k \gamma_5 b_{\partial_i}$                              &$+$\\[0.3cm]
14    &  $\epsilon_{ijk}\ov{a}_{\partial_k} \gamma_j b_n+\epsilon_{ijk}\ov{a}_n \gamma_j b_{\partial_k}$         &$-$\\
15    &  $\epsilon_{ijk}\ov{a}_{\partial_k} \gamma_j b_n-\epsilon_{ijk}\ov{a}_n \gamma_j b_{\partial_k}$         &$+$\\
16    &  $\epsilon_{ijk}\ov{a}_{\partial_k} \gamma_j b_w+\epsilon_{ijk}\ov{a}_w \gamma_j b_{\partial_k}$         &$-$\\
17    &  $\epsilon_{ijk}\ov{a}_{\partial_k} \gamma_j b_w-\epsilon_{ijk}\ov{a}_w \gamma_j b_{\partial_k}$         &$+$\\[0.3cm]
18    &  $\epsilon_{ijk}\ov{a}_{\partial_k} \gamma_j\gamma_t b_n+\epsilon_{ijk}\ov{a}_n \gamma_j\gamma_t b_{\partial_k}$ &$-$\\
19    &  $\epsilon_{ijk}\ov{a}_{\partial_k} \gamma_j\gamma_t b_n-\epsilon_{ijk}\ov{a}_n \gamma_j\gamma_t b_{\partial_k}$ &$+$\\
20    &  $\epsilon_{ijk}\ov{a}_{\partial_k} \gamma_j\gamma_t b_w+\epsilon_{ijk}\ov{a}_w \gamma_j\gamma_t b_{\partial_k}$ &$-$\\
21    &  $\epsilon_{ijk}\ov{a}_{\partial_k} \gamma_j\gamma_t b_w-\epsilon_{ijk}\ov{a}_w \gamma_j\gamma_t b_{\partial_k}$ &$+$\\[0.3cm]
22    &  $\ov{a}_n \gamma_k\gamma_t\gamma_5 b_n$                                     &$-$\\
23    &  $\ov{a}_n \gamma_k\gamma_t\gamma_5 b_w+\ov{a}_w \gamma_k\gamma_t\gamma_5 b_n$                  &$-$\\
24    &  $\ov{a}_n \gamma_k\gamma_t\gamma_5 b_w-\ov{a}_w \gamma_k\gamma_t\gamma_5 b_n$                  &$+$\\
25    &  $\ov{a}_w \gamma_k\gamma_t\gamma_5 b_w$                                     &$-$\\[0.3cm]
26    &  $\ov{a}_{\partial_i} \gamma_k\gamma_t\gamma_5 b_{\partial_i}$                          &$-$\\
\end{tabular}
\end{ruledtabular}
\caption{Same as Tab.\ \ref{tab:interpols_0-}, now for $J^P=1^+$.}
\label{tab:interpols_1+}
\end{table}

\begin{table}[!]
\begin{ruledtabular}
\begin{tabular}{lcc}
\#$_{2^-E}$ & interpolator(s) & $C$\\
\hline
1  &  $Q_{ijk} \bar{a}_{\partial_k} \gamma_j\gamma_t\gamma_5 b_n+Q_{ijk} \bar{a}_n \gamma_j\gamma_t\gamma_5 b_{\partial_k}$& $-$ \\
2  &  $Q_{ijk} \bar{a}_{\partial_k} \gamma_j\gamma_t\gamma_5 b_n-Q_{ijk} \bar{a}_n \gamma_j\gamma_t\gamma_5 b_{\partial_k}$& $+$ \\
3  &  $Q_{ijk} \bar{a}_{\partial_k} \gamma_j\gamma_t\gamma_5 b_w+Q_{ijk} \bar{a}_w \gamma_j\gamma_t\gamma_5 b_{\partial_k}$& $-$ \\
4  &  $Q_{ijk} \bar{a}_{\partial_k} \gamma_j\gamma_t\gamma_5 b_w-Q_{ijk} \bar{a}_w \gamma_j\gamma_t\gamma_5 b_{\partial_k}$& $+$ \\[0.3cm]
5  &  $Q_{ijk} \bar{a}_{\partial_j} \gamma_5 b_{\partial_k}$& $+$ \\[0.3cm]
6  &  $Q_{ijk} \bar{a}_{\partial_j} \gamma_t\gamma_5 b_{\partial_k}$& $+$ \\[0.3cm]
7  &  $Q_{ijk} \bar{a}_{\partial_k} \gamma_j\gamma_5 b_n+Q_{ijk} \bar{a}_n \gamma_j\gamma_5 b_{\partial_k}$& $+$ \\
8  &  $Q_{ijk} \bar{a}_{\partial_k} \gamma_j\gamma_5 b_n-Q_{ijk} \bar{a}_n \gamma_j\gamma_5 b_{\partial_k}$& $-$ \\
9  &  $Q_{ijk} \bar{a}_{\partial_k} \gamma_j\gamma_5 b_w+Q_{ijk} \bar{a}_w \gamma_j\gamma_5 b_{\partial_k}$& $+$ \\
10  &  $Q_{ijk} \bar{a}_{\partial_k} \gamma_j\gamma_5 b_w-Q_{ijk} \bar{a}_w \gamma_j\gamma_5 b_{\partial_k}$& $-$ \\
\end{tabular}
\end{ruledtabular}
\caption{Same as Tab.\ \ref{tab:interpols_0-}, now for $J^P=2^-E$.}
\end{table}

\begin{table}[!]
\begin{ruledtabular}
\begin{tabular}{lcc}
\#$_{2^+E}$ & interpolator(s) & $C$\\
\hline
1  &   $Q_{ijk} \bar{a}_{\partial_k} \gamma_j b_n+Q_{ijk} \bar{a}_n \gamma_j b_{\partial_k}$         &$-$\\
2  &   $Q_{ijk} \bar{a}_{\partial_k} \gamma_j b_n-Q_{ijk} \bar{a}_n \gamma_j b_{\partial_k}$         &$+$\\
3  &   $Q_{ijk} \bar{a}_{\partial_k} \gamma_j b_w+Q_{ijk} \bar{a}_w \gamma_j b_{\partial_k}$         &$-$\\
4  &   $Q_{ijk} \bar{a}_{\partial_k} \gamma_j b_w-Q_{ijk} \bar{a}_w \gamma_j b_{\partial_k}$         &$+$\\[0.3cm]
5  &   $Q_{ijk} \bar{a}_{\partial_k} \gamma_j\gamma_t b_n+Q_{ijk} \bar{a}_n \gamma_j\gamma_t b_{\partial_k}$ &$-$\\
6  &   $Q_{ijk} \bar{a}_{\partial_k} \gamma_j\gamma_t b_n-Q_{ijk} \bar{a}_n \gamma_j\gamma_t b_{\partial_k}$ &$+$\\
7  &   $Q_{ijk} \bar{a}_{\partial_k} \gamma_j\gamma_t b_w+Q_{ijk} \bar{a}_w \gamma_j\gamma_t b_{\partial_k}$ &$-$\\
8  &   $Q_{ijk} \bar{a}_{\partial_k} \gamma_j\gamma_t b_w-Q_{ijk} \bar{a}_w \gamma_j\gamma_t b_{\partial_k}$ &$+$\\[0.3cm]
9  &   $Q_{ijk} \bar{a}_{\partial_j} b_{\partial_k}$                             &$+$\\[0.3cm]
10  &   $Q_{ijk} \bar{a}_{\partial_j} \gamma_t b_{\partial k}$                         &$-$\\
\end{tabular}
\end{ruledtabular}
\caption{Same as Tab.\ \ref{tab:interpols_0-}, now for $J^P=2^+E$.}
\label{tab:interpols_2+E}
\end{table}

\begin{table}[!]
\begin{ruledtabular}
\begin{tabular}{lcc}
\#$_{2^-T_2}$ & interpolator(s) & $C$\\
\hline
1  &  $|\epsilon_{ijk}| \bar{a}_{\partial_k} \gamma_j\gamma_5 b_n+|\epsilon_{ijk}| \bar{a}_n \gamma_j\gamma_5 b_{\partial_k}$         &$+$\\
2  &  $|\epsilon_{ijk}| \bar{a}_{\partial_k} \gamma_j\gamma_5 b_n-|\epsilon_{ijk}| \bar{a}_n \gamma_j\gamma_5 b_{\partial_k}$         &$-$\\
3  &  $|\epsilon_{ijk}| \bar{a}_{\partial_k} \gamma_j\gamma_5 b_w+|\epsilon_{ijk}| \bar{a}_w \gamma_j\gamma_5 b_{\partial_k}$         &$+$\\
4  &  $|\epsilon_{ijk}| \bar{a}_{\partial_k} \gamma_j\gamma_5 b_w-|\epsilon_{ijk}| \bar{a}_w \gamma_j\gamma_5 b_{\partial_k}$         &$-$\\[0.3cm]
5  &  $|\epsilon_{ijk}| \bar{a}_{\partial_k} \gamma_j\gamma_t\gamma_5 b_n+|\epsilon_{ijk}| \bar{a}_n \gamma_j\gamma_t\gamma_5 b_{\partial_k}$ &$-$\\
6  &  $|\epsilon_{ijk}| \bar{a}_{\partial_k} \gamma_j\gamma_t\gamma_5 b_n-|\epsilon_{ijk}| \bar{a}_n \gamma_j\gamma_t\gamma_5 b_{\partial_k}$ &$+$\\
7  &  $|\epsilon_{ijk}| \bar{a}_{\partial_k} \gamma_j\gamma_t\gamma_5 b_w+|\epsilon_{ijk}| \bar{a}_w \gamma_j\gamma_t\gamma_5 b_{\partial_k}$ &$-$\\
8  &  $|\epsilon_{ijk}| \bar{a}_{\partial_k} \gamma_j\gamma_t\gamma_5 b_w-|\epsilon_{ijk}| \bar{a}_w \gamma_j\gamma_t\gamma_5 b_{\partial_k}$ &$+$\\
\end{tabular}
\end{ruledtabular}
\caption{Same as Tab.\ \ref{tab:interpols_0-}, now for $J^P=2^-T_2$.}
\end{table}

\begin{table}[!]
\begin{ruledtabular}
\begin{tabular}{lcc}
\#$_{2^+T_2}$ & interpolator(s) & $C$\\
\hline
1  &  $|\epsilon_{ijk}| \bar{a}_{\partial_k} \gamma_j b_n+|\epsilon_{ijk}| \bar{a}_n \gamma_j b_{\partial_k}$         &$-$\\
2  &  $|\epsilon_{ijk}| \bar{a}_{\partial_k} \gamma_j b_n-|\epsilon_{ijk}| \bar{a}_n \gamma_j b_{\partial_k}$         &$+$\\
3  &  $|\epsilon_{ijk}| \bar{a}_{\partial_k} \gamma_j b_w+|\epsilon_{ijk}| \bar{a}_w \gamma_j b_{\partial_k}$         &$-$\\
4  &  $|\epsilon_{ijk}| \bar{a}_{\partial_k} \gamma_j b_w-|\epsilon_{ijk}| \bar{a}_w \gamma_j b_{\partial_k}$         &$+$\\[0.3cm]
5  &  $|\epsilon_{ijk}| \bar{a}_{\partial_k} \gamma_j\gamma_t b_n+|\epsilon_{ijk}| \bar{a}_n \gamma_j\gamma_t b_{\partial_k}$ &$-$\\
6  &  $|\epsilon_{ijk}| \bar{a}_{\partial_k} \gamma_j\gamma_t b_n-|\epsilon_{ijk}| \bar{a}_n \gamma_j\gamma_t b_{\partial_k}$ &$+$\\
7  &  $|\epsilon_{ijk}| \bar{a}_{\partial_k} \gamma_j\gamma_t b_w+|\epsilon_{ijk}| \bar{a}_w \gamma_j\gamma_t b_{\partial_k}$ &$-$\\
8  &  $|\epsilon_{ijk}| \bar{a}_{\partial_k} \gamma_j\gamma_t b_w-|\epsilon_{ijk}| \bar{a}_w \gamma_j\gamma_t b_{\partial_k}$ &$+$\\
\end{tabular}
\end{ruledtabular}
\caption{Same as Tab.\ \ref{tab:interpols_0-}, now for $J^P=2^+T_2$.}
\label{tab:interpols_2+T_2}
\end{table}

\begin{table}[!]
\begin{ruledtabular}
\begin{tabular}{lcc}
light meson 	& energy level [MeV]	& $\chi^2$/d.o.f. 	\\
\hline
$0^{-+}$		& 1407(103)		& 8.25/5		\\
$0^{++}$		& 976(70)		& 12.38/5		\\
$0^{++}$		& 1689(103)		& 6.70/4		\\
$1^{--}$		& 772(13)		& 4.65/5		\\
$1^{--}$		& 1528(115)		& 2.79/5		\\
$1^{--}$		& 1733(143)		& 2.51/4		\\
$1^{-+}$		& 1370(260)		& 3.78/5		\\
$1^{+-}$		& 1347(26)		& 8.12/5		\\
$1^{++}$		& 1238(33)		& 9.62/5		\\
$1^{++}$		& 1754(103)		& 4.91/5		\\
$2^{--}(T_2)$	& 1849(222)		& 3.77/2		\\
$2^{--}(E)$		& 1965(183)		& 2.16/3		\\
$2^{-+}(T_2)$	& 1745(96)		& 5.06/5		\\
$2^{-+}(E)$		& 1889(139)		& 8.96/4		\\
$2^{++}(T_2)$	& 1399(66)		& 16.51/5		\\
$2^{++}(E)$		& 1379(60)		& 6.03/5		\\
\end{tabular}
\end{ruledtabular}
\caption{Energy levels at the physical point and corresponding $\chi^2$/d.o.f.~for the chiral fits of the isovector light meson energy levels reported in this work.
Sources of large $\chi^2$/d.o.f.~($\geq 3$) are discussed in the text.}
\label{tab:chi2lightmesons}
\end{table}

\begin{table}[!]
\begin{ruledtabular}
\begin{tabular}{lcc}
strange meson 	& energy level [MeV]	& $\chi^2$/d.o.f. 	\\
\hline
$0^{-}$		& 509(4)		& 20.83/5		\\
$0^{-}$		& 1434(64)		& 6.94/5		\\
$0^{+}$		& 884(36)		& 41.53/5		\\
$0^{+}$		& 1323(81)		& 6.92/5		\\
$1^{-}$		& 896(9)		& 7.20/5		\\
$1^{-}$		& 1633(89)		& 1.57/3		\\
$1^{-}$		& 1919(69)		& 1.05/3		\\
$1^{+}$		& 1339(20)		& 3.90/5		\\
$1^{+}$		& 1409(17)		& 6.76/5		\\
$1^{+}$		& 1709(109)		& 2.56/5		\\
$2^{-}(T_2)$	& 1750(54)		& 5.31/5		\\
$2^{-}(T_2)$	& 1909(52)		& 2.21/5		\\
$2^{-}(E)$		& 1870(75)		& 1.51/4		\\
$2^{-}(E)$		& 1956(71)		& 1.17/4		\\
$2^{+}(T_2)$	& 1452(51)		& 7.28/5		\\
$2^{+}(E)$		& 1392(58)		& 5.68/5		\\
\end{tabular}
\end{ruledtabular}
\caption{Same as Tab.~\ref{tab:chi2lightmesons}, but for strange mesons.}
\label{tab:chi2strangemesons}
\end{table}

\begin{table}[!]
\begin{ruledtabular}
\begin{tabular}{lcc}
isoscalar meson 	& energy level [MeV]	& $\chi^2$/d.o.f. 	\\
\hline
$1^{--}$		& 994(8)		& 6.51/5		\\
$1^{--}$		& 1857(53)		& 7.30/5		\\
$1^{--}$		& 1987(40)		& 1.41/5		\\
$2^{++}(T_2)$		& 1581(29)		& 12.89/5		\\
$2^{++}(E)$		& 1578(24)		& 7.28/5		\\
\end{tabular}
\end{ruledtabular}
\caption{Same as Tab.~\ref{tab:chi2lightmesons}, but for isoscalar mesons.}
\label{tab:chi2isoscalarmesons}
\end{table}

\end{appendix}

\clearpage
\bibliographystyle{apsrev4-1}
%

\end{document}